\documentclass[12pt,awide]{article}
\usepackage{latexsym,a4,a4wide,psfrag,here,
amssymb,amsmath,axodraw,color,ifthen,mathrsfs}
\usepackage[dvips]{graphicx}
\usepackage[mathcal]{eucal}

\def\mA{\ensuremath{\mathcal{A}}}

\def\mD{\ensuremath{\mathcal{D}}}

\def\mG{\ensuremath{\mathcal{G}}}

\def\mL{\ensuremath{\mathcal{L}}}

\def\mO{\ensuremath{\mathcal{O}}}
\def\mQ{\ensuremath{\mathcal{Q}}}

\def\wvb{\ensuremath{\bar{\mathfrak{w}}}}
\def\svb{\ensuremath{\bar{\mathfrak{s}}}}

\def\nn{\nonumber \\}
\def\no{\nonumber}
\def\nnn{ \nonumber \\ & & }
\def\fs{\; \; .}
\def\co{\; \; ,}
\def\con{\;\;, \nonumber \\}
\def\sem{\; \; ;}
\def\fs{\; \; .}
\def\co{\; \; ,}
\def\sem{\; \; ;}

\def\mexp{\mbox{exp}}
\def\mln{\mbox{ln}}

\def\mTr{\mbox{Tr}}

\def\Deb{\ensuremath{\bar{\Delta}}}

\def\vph{\ensuremath{\varphi}}

\def\eps{\ensuremath{\varepsilon}}

\def\don{_{\nu}}
\def\dom{_{\mu}}
\def\domn{_{\mu\nu}}

\def\upmn{^{\mu\nu}}

\def\pa{\partial}

\def\ddm{\partial_\mu}
\def\ddn{\partial_\nu}

\def\lai{\lambda_{i}}
\def\laj{\lambda_{j}}

\def\Dm{\ensuremath{\nabla_{\mu}}}
\def\Dn{\ensuremath{\nabla_{\nu}}}

\def\cpt{\mbox{CHPT\,}}

\def\bdm{\begin{displaymath}}
\def\edm{\end{displaymath}}
\def\be{\begin{equation}}
\def\ee{\end{equation}}

\def\bea{\begin{eqnarray}}
\def\eea{\end{eqnarray}}
\def\bes{\begin{eqnarray*}}
\def\ees{\end{eqnarray*}}

\def\ba{\begin{array}}
\def\ea{\end{array}}
\def\bal{\begin{align}}
\def\eal{\end{align}}
\def\bspl{\begin{split}}
\def\espl{\end{split}}
\def\bml{\begin{multline}}
\def\eml{\end{multline}}

\def\um{\ensuremath{u_{\mu}}}
\def\umb{\ensuremath{\bar{u}_{\mu}}}
\def\uum{\ensuremath{u^{\mu}}}

\def\un{\ensuremath{u_{\nu}}}
\def\unb{\ensuremath{\bar{u}_{\nu}}}

\def\xim{\ensuremath{\xi_{\mu}}}
\def\xin{\ensuremath{\xi_{\nu}}}

\def\De{\ensuremath{\Delta}}
\def\Dem{\ensuremath{\Delta_{\mu}}}
\def\Demb{\ensuremath{\bar{\Delta}_{\mu}}}

\def\cpl#1#2{\chi_{_{+}#1}^{\;\;\,#2}}
\def\cml#1#2{\chi_{_{-}#1}^{\;\;\,#2}}
\def\ph{\;\;}

\def\cp{\ensuremath{\chi_{_{+}}}}

\def\cpb{\ensuremath{\bar{\chi}_{_{+}}}}

\def\cm{\ensuremath{\chi_{_{-}}}}

\def\cipm{\ensuremath{\chi_{_{\pm}}}}
\def\cipmb{\ensuremath{\bar{\chi}_{_{\pm}}}}

\def\cimpb{\ensuremath{\bar{\chi}_{_{\mp}}}}

\def\dcihpm{\ensuremath{\chi_{_{\pm , \,\mu}}}}
\def\dcihpmb{\ensuremath{\bar{\chi}_{_{\pm , \,\mu}}}}

\def\dcihmpb{\ensuremath{\bar{\chi}_{_{\mp , \,\mu}}}}

\def\cmb{\ensuremath{\bar{\chi}_{_{-}}}}
\def\cmh{\ensuremath{\hat{\chi}_{_{-}}}}

\def\usq{\ensuremath{u^{2}}}
\def\usqb{\ensuremath{\bar{u}^{2}}}

\def\fpdmn{\ensuremath{f_{ + \mu \nu}}}

\def\fipm{\ensuremath{f_{\pm}\upmn}}
\def\fipmb{\ensuremath{\bar{f}_{\pm}\upmn}}

\def\fimpb{\ensuremath{\bar{f}_{\mp}\upmn}}

\def\hmn{\ensuremath{h\domn}}

\def\umn{\ensuremath{u_{\mu\nu}}}
\def\umnb{\ensuremath{\bar{u}_{\mu\nu}}}

\def\ltr{\langle}
\def\rtr{\rangle}

\def\ktppp{\ensuremath{K \to \pi \pi \pi \,\,}}
\def\ktpp{\ensuremath{K \to \pi \pi \,\,}}

\def\bdm{\begin{displaymath}}
\def\edm{\end{displaymath}}

\renewcommand{\theequation}{\arabic{equation}}

\def\GeV{\mbox{GeV}}
\def\MeV{\mbox{MeV}}

\def\ho{$ \hbar $-order }

\def\xh{\ensuremath{\hat{X}}}

\definecolor{gray}{gray}{.5}
\definecolor{steelblue}{rgb}{0.153, 0.102, 0.255}
\definecolor{lightblue}{rgb}{0,0.2,0.5}

\renewcommand{\i}{{\rm i}}

\renewcommand{\theequation}{\arabic{section}.\arabic{equation}}

\begin{document}
\bibliographystyle{aipep}
 \title{The leading two loop divergences of the nonleptonic weak chiral Lagrangian}
\author{
 M.~B\"uchler\\
{\small
Department of Physics, University of Washington}\\
{\small Seattle, WA 98195-1560, U.S.A. }}
\maketitle
\begin{abstract}
\noindent We calculate the leading divergences at NNLO
for the octet part of the nonleptonic weak sector of chiral perturbation theory, using renormalization
group methods. The role of counterterms which vanish at
the equation of motion and their use to simplify the calculation is shown explicitly.
The obtained counterterm Lagrangian can be employed to calculate the chiral double log
contributions of quantities in this sector, most notably the \ktpp amplitude.
The double log contribution of the latter is discussed in a separate paper. 
\end{abstract}
\newpage
\section{Introduction}
We determine the leading divergences at NNLO for the nonleptonic weak chiral Lagrangian which 
transforms as an octet under the chiral group, extending the NLO calculation of the latter
\cite{Kambor:1990tz,Ecker:1993de}.  
The obtained counterterm Lagrangian can be used to calculate the leading logarithmic contributions, double logs in short, of observables in this sector.
These contributions are in particular interesting since all the low energy constants of higher order are unknown in this sector. 
The calculation of logarithmic contributions provides thus the only way to get a 
first analytical estimate of the size of the higher order corrections one has to expect. Analogue calculations in the strong sector of chiral perturbation theory (\cpt) have already been worked out \cite{Bijnens:1998yu,Bijnens:1999hw}.
The results of this paper are used to calculate the double logs
to the \ktpp amplitude, presented in a separate paper \cite{dlkspp}.
For the latter amplitude, methods have also been worked out to extract the needed
next to leading order (NLO) low energy constants (LEC's) by lattice simulations;
however, the proposed approach is rather ambitious \cite{Boucaud:2001mg,Laiho:2002jq}.
Interesting further applications of the results obtained here are for instance the calculation of the double logs of the \ktppp or the  $K \to \pi \gamma \gamma$  amplitude. \\

\noindent The chiral logs are introduced during the process of renormalization
\cite{Li:1971vr}. These logarithms, which correspond to the infrared singularities
when the masses of the theory approach zero, can produce sizable contributions
to observables. 
\noindent Using dimensional regularization, it is straightforward to understand how the leading logs 
are related to the leading counterterms: while renormalizing the theory, 
one has to introduce an energy scale $\mu$ to
ensure the correct dimensions of observables calculated in $d-$dimensional space-time.
In particular, for the divergences generated by loop calculations, this means that
they can only show up in the following structures:
\bea
 \mQ & := & \frac{\mu^{d-4}}{(4\pi)^{d/2}} \Big( \frac{1}{4-d} - \frac{1}{2} \mbox{ln}(\frac{m^{2}}{\mu^{2}}) \Big)
 \fs
 \label{eq:qs}
\eea

\noindent To illustrate the order of magnitudes of these chiral corrections,
Table \ref{t:bl} displays the various contributions 
up to NNLO for the $\pi\pi$ scattering lengths in two flavor \cpt, showing 
that the double log contribution in this case amounts to almost the full NNLO corrections, 
corresponding to close to $10\%$ of the total result \cite{Bijnens:1995yn}: \\
\begin{table*}[h]
\begin{center}
\begin{tabular}{ccccc}
 & LO & NLO & NNLO & DOUBLE LOGS\\
 \hline
 $a^{0}_{0}$ & 0.156 & 0.044 & 0.017 & 0.013\\
 $a^{0}_{0}-a^{2}_{0}$ & 0.201 & 0.042 & 0.016 & 0.012\\
\end{tabular}
\end{center}
\caption{The leading order $\pi\pi$ scattering lengths in two flavor \cpt with the chiral corrections up to NNLO, for some standard values of the NLO LEC's and at a renormalization scale $\mu = 1\GeV$. The double logs are included in the NNLO correction.}
\label{t:bl}
\end{table*}

\noindent For the case of three flavor \cpt, we
 show the chiral corrections up to NNLO to the pion and Kaon decay constants and the vector form factor of $K_{l3}$ \cite{Amoros:1999dp,Bijnens:2003uy,Bijnens:1998yu} in Table \ref{t:b2l}.
One notices that the relative size of the double logs is less pronounced than for two flavor \cpt. Typically, the double logs amount to $20 - 35 \%$ of the total NNLO contributions,
corresponding to about $10\%$ of the total corrections to the leading order result.
Although the numerical values of these corrections are not too large, one should keep in mind that for applications like chiral extrapolations, the relative size of the double log contributions to the NLO corrections is of importance, which is in the range of $10\%$ and
therefore sizable. Let us in this context again emphasize that while it is certainly appropriate to perform the full two loop calculations in
the strong sector, due to the lack of knowledge of the weak LEC's  the calculation of the double log contributions is presumably the best one can ever achieve to get an estimate of the size of the NNLO corrections in the weak sector.\\
\begin{table*}[h]
\begin{center}
\begin{tabular}{ccccc}
 & LO & NLO & NNLO & DOUBLE LOGS \\
 \hline
 $F_{\pi}/F_{0}$ & 1 & 0.068 & -0.172 & -0.050 \\
 $F_{K}/F_{\pi}$ & 1 & 0.216 & 0.035 & 0.06 \\
 $f_{+}(0)[K_{l3}]$ & 1 & -0.023 & 0.015 & 0.004 \\
\end{tabular}
\end{center}
\caption{The chiral corrections up to NNLO for the pion and Kaon decay constants and the
vector form factor of $K_{l3}$.
The values can however vary considerably, depending on the LEC's one employs. The numbers above are calculated with some standard values of the NLO LEC's and all
the renormalized NNLO LEC's set to zero at $\mu=770\MeV$.}
\label{t:b2l}
\end{table*}
\\
\noindent The outline of the paper is as follows: \\
In section \ref{s:cpt} we introduce the notation used and  give the needed \cpt Lagrangians.
Section \ref{s:gfnwl} provides a very brief overview of  the
general framework in which this calculation was performed; the generating functional is introduced, and it is shown how one can use the background field method to calculate the counterterms needed to renormalize the latter. \\
In section \ref{s:eomt} we discuss the role of operators which vanish at the solution of
the equation of motion (EOM terms) to simplify the calculation of counterterm Lagrangians:
One can choose the coefficients of these EOM terms in a way that the sum of all one particle 
reducible (1PR) topologies at a given \ho won't generate any divergences. We pin down these 
coefficients and henceforth only need to take into account one particle irreducible (1PI) topologies. \\
In section \ref{s:rge}, we sketch very briefly the renormalization group techniques which are
employed in the present calculation. The basic result of this section is that one can 
obtain the \ho 2 highest pole counterterm (NNLO) by performing a one loop calculation which uses 
the \ho 1 (NLO) counterterm as input.
In the last part, section \ref{s:oc}, we illustrate how the concrete calculation works with two simple examples.
\setcounter{equation}{0}
\section{\cpt Lagrangian}
\label{s:cpt}
The lowest order chiral Lagrangian which allows for $\De S =1 $ strangeness changing interactions
is given by (Throughout this section we will work in euclidean space-time):
\be
 \mL^{(0)} = \mL^{(0)}_{\tiny{s}} + \mL^{(0)}_{\tiny{\De S = 1}} \co
 \label{eq:slocl}
\ee
which encodes the dynamics of the pseudo-Goldstone bosons in the presence of external
source fields $s,p,v_{\mu},a_{\mu}$. The first term corresponds to the strong interaction
Lagrangian:
\be
 \mL^{(0)}_{\tiny{s}} = \frac{F^{2}_{0}}{4} \big( \ltr \um\um \rtr - \ltr \cp \rtr \big) \co
 \label{eq:loscl}
\ee
where $\ltr \cdot \rtr$ stands for the flavor trace. Further we used:
\bea
 \um & = & \i \big( u^{\dagger}(\pa\dom - \i r\dom )u - u(\pa\dom - \i l\dom) u^{\dagger} \big) \con \nn
\cp & = & u^{\dagger} \chi u^{\dagger} + u \chi^{\dagger} u \fs
\label{eq:loscld}
\eea
A list of additional building blocks used for the Lagrangians of higher order can be found in appendix \ref{s:nd}. \\
The $u$ matrix encodes the octet of the
light pseudo-scalar bosons in the exponential parametrization:
\be
u  = \exp(\frac{\i \phi}{\sqrt{2} F})\quad ; \quad  \phi =
\left(
\begin{array}{c c c}
 \frac{\pi^{0}}{\sqrt{2}} + \frac{\eta_{8}}{\sqrt{6}} & \pi^{+} & K^{+} \\
 \pi^{-} & -\frac{\pi^{0}}{\sqrt{2}} + \frac{\eta_{8}}{\sqrt{6}} & K^{0} \\
 K^{-}  & \bar{K}^{0} & -\frac{2 \eta_{8}}{\sqrt{6}}
\end{array}
\right) \fs
\label{eq:phid}
\ee
The definitions in Eq. (\ref{eq:loscld}) embody also $r\dom$ and
$l\dom$, the external vector source fields, whereas the scalar counterparts 
are encoded in $\chi$
($\mathscr{M}$ being the quark mass matrix):
\be
 \chi = 2 B_{0}\big( \mathscr{M} + s(x) + \i p(x) \big) \fs \no
\ee
$B_{0}$ is related to the vacuum expectation value of the scalar quark density: 
\be
 \ltr 0 | \bar{q} q | 0 \rtr = - F_{0}^{2}B_{0}(1 + \mO(\mathscr{M})) \fs \no  
\ee
$F_{0} \simeq 92.4 \MeV$ corresponds to the chiral limit value of the pion decay constant. \\
The Lagrangian which triggers flavor changing processes is given by: 
\bea 
 \mL^{(0)}_{\tiny{\De S = 1}} & = &C F_{0}^{4} \Big( g_{8} \ltr \De_{32} \um \um \rtr - g_{8}^{'} \ltr \De_{32} \cp \rtr + g_{27} t^{ij;kl} \ltr \De_{ij} \um \rtr \ltr \De_{kl} \um \rtr \Big) + \mbox{h.c.} \co 
 \label{eq:elowcl}
\eea 
\\
with $t^{11;23} = t^{13;21}=t^{21;13}=t^{23;11}=1/3\; , \; t^{22;23} = t^{23;22} =t^{23;33}=t^{33;23} = -1/6 $, and all other $t$'s vanishing. \\
The constant $C$:
\be
 C = -\frac{G_{F}}{\sqrt{2}}V_{ud}V_{us}^{*} \co 
\ee
renders the coupling constants $g_{8},g_{8}^{'}$ and $g_{27}$ dimensionless. \\
$\De_{ij}$ is defined as:
\begin{alignat}{2}
 \De_{ij} & := u \lambda_{ij} u^{\dagger} \quad ; \quad & 
 (\lambda_{ij})_{ab} & = \delta_{ai}\delta_{bj} \fs \no
\end{alignat}
The first two operators in Eq. (\ref{eq:elowcl}), proportional to $g_{8}$ and $g_{8}^{'}$ transform as an octet, $(8,1)$,
under the chiral group $\mbox{SU}(3)_{L}\otimes \mbox{SU}(3)_{R}$, whereas the last, 
proportional to $g_{27}$, transforms like a $27$-plet, $(27,1)$.
We neglected a third contribution transforming like $(8,8)$, which takes into account virtual 
photons.  \\
Since the octet part of Eq. (\ref{eq:elowcl}) is believed to be the main source of the
$\De I = 1/2$ rule, we will only use the latter in our calculation. Furthermore we discard
the part proportional to $g_{8}^{'}$, since it doesn't contribute to on-shell processes
\cite{Bernard:1985wf,Crewther:1985zt,Leurer:1987ih,Kambor:1990tz}.
Of the remaining operator we only use its CP even part, which we will henceforth denote
by $\mL^{(0)}_{\tiny{w}}$:
\begin{alignat}{2}
 \mL^{(0)}_{\tiny{w}} & := C F_{0}^{4} g_{8} \ltr \De \um \uum \rtr \quad ; &
  \quad \De := u \lambda_{6} u^{\dagger} \fs
  \label{eq:lowcl}
\end{alignat}
In addition to the lowest order Lagrangians $\mL^{(0)}_{\tiny{s}}$ and $\mL^{(0)}_{\tiny{w}}$ discussed above, we will also use the NLO Lagrangian $\mL^{(1)}_{\tiny{w}}$, introduced in Eq. (\ref{eq:wcl1}).
\setcounter{equation}{0}
\section{The $\hbar$ expansion of the generating functional}
\label{s:gfnwl}
In this section we set up some notation and discuss the NLO and NNLO expressions
for the generating functional of the nonleptonic weak chiral Lagrangian.
\subsection{Notation}
The generating functional is defined as the vacuum to vacuum transition amplitude in the
presence of sources, collectively denoted by $j$ (Throughout this section we will
be working in euclidean space-time.):
\be
\label{eq:gf}
 e^{Z[j]/\hbar} = N \int \Pi[\mD\vph_{i}] e^{-S[\vph,j]/\hbar} \fs
\ee
$Z[j]$ as well as $S[\vph,j]$ can be split into a strong and weak part:
\begin{alignat}{2}
 Z[j] & = Z_{\tiny{s}}[j] + Z_{\tiny{w}}[j]\quad ; \quad &  S[\vph,j] & = S^{\tiny{s}}[\vph,j] + S^{\tiny{w}}[\vph,j]  \co
\end{alignat}
analogously to Eq. (\ref{eq:slocl}). These can be expanded in their \ho:
\begin{alignat}{2}
 Z_{\tiny{s}}[j] = \sum_{n=0}^{\infty} Z_{\tiny{s}}^{(n)}[j] \quad ; \quad &
 S^{\tiny{s}}[\vph,j] = \sum_{n}^{\infty} S^{\tiny{s}}_{n}[\vph,j]  \co 
\end{alignat}
and similarly for the weak part. \\
The tree level generating functional $Z^{(0)}[j]$ corresponds to the action with the lowest order chiral Lagrangian, evaluated at the EOM: 
\be 
Z^{(0)}[j] = S^{(0)}[\bar{\vph},j] \fs 
\ee
One can calculate the separate \ho contributions to the generating functional by use of the background field method, which is briefly outlined in appendix \ref{app:bfm}: 
The strong and weak nonleptonic
 chiral actions are expanded in quantum fluctuation fields $\xi$ around a field $\bar{\vph}$, which in the following is assumed to be a solution of the classical equation of motion, i.e. $\svb^{i}_{0} = 0$. To keep the notation simple, we will henceforth suppress the arguments of $Z[j]$, $S[\vph,j]$ and $\mL[\vph]$:
\bea
 S^{\tiny{s}} & = & \bar{S}^{\tiny{s}} + \svb^{i}\xi_{i}
 + \frac{1}{2!}\svb^{ij}\xi_{i}\xi_{j}
 + \frac{1}{3!}\svb^{ij}\xi_{i}\xi_{j}\xi_{k}
 + \frac{1}{4!}\svb^{ijkl}\xi_{i}\xi_{j}\xi_{k}\xi_{l} + \mO(\xi^{5})
 \con
 S^{\tiny{w}} & = & \bar{S}^{\tiny{w}} + \wvb^{i}\xi_{i}
 + \frac{1}{2!}\wvb^{ij}\xi_{i} \xi_{j}
  + \frac{1}{3!}\wvb^{ijk}\xi_{i}\xi_{j}\xi_{k}
 + \frac{1}{4!}\wvb^{ijkl}\xi_{i}\xi_{j}\xi_{k}\xi_{l} + \mO(\xi^{5})
 \fs \no
\eea
This expansion provides the vertices which will be needed for the calculation of $Z^{(2)}_{\tiny{w}}$, Eq. (\ref{eq:z2}). The Latin indices $i,j,..$ correspond to an $SU(N)$ index as well as a space-time 
degree of freedom. The field $\bar{\vph}$ is treated as a background field, and the $\xi$-field is employed as new integration variable in the path integral, Eq. (\ref{eq:gf}).
The perturbative evaluation of this "new" generating functional results in  vacuum diagrams with respect to the $\xi$ fields. 
\subsection{NLO: $Z^{(1)}_{\tiny{w}}$}
The counterterms needed to renormalize  $Z^{(1)}_{\tiny{w}}$:
 \bea
  Z^{(1)}_{\tiny{w}} & = &
  \frac{1}{2}  \wvb^{ i j }_{0} G_{ij} 
  + \bar{S}^{\tiny{w}}_{1} \co
  \label{eq:wz1}
 \eea
were first calculated by Kambor, Missimer and Wyler \cite{Kambor:1990tz}. Throughout this paper, we will however use
the basis of operators given by Ecker, Kambor and Wyler (EKW) \cite{Ecker:1993de},
who used the EOM to reduce the former. This Lagrangian assumes the form:
\be
 \mL_{\tiny{w}} ^{(1)} = C g_{8}F^{2}_{0}\sum_{i=1}^{37} c_{i}^{(1)}W_{i}^{(1)} \co
 \label{eq:wcl1}
\ee
where the bare LEC's $c_{i}^{(1)}$ are split into a renormalized and counterterm part:
\be
 c_{i}^{(1)} = (\mu c)^{-\eps} \big( c_{i}^{(1) r}(\mu,\eps) +  a_{1\,i}^{(1)} \Lambda \big) \fs
 \label{eq:wlec1}
\ee
$\mu$ is the renormalization scale and the constant $c$ parametrizes the regularization prescription ( $\mln(c) =  \exp(-( \mln(4\pi) + \Gamma^{'}(1) + 1 )/2$  for $\overline{\mbox{MS}}$ ).)
In addition we use the notation:
\begin{alignat}{3}
 \eps & := 4-d \sem \quad & \hat{N} & := (4\pi)^{-2} \sem \quad & \Lambda  & :=   \frac{\hat{N}}{\eps}  \fs
\end{alignat}
In Table \ref{t:wc}, we list the operators of the EKW basis needed for the \ktpp amplitude at NLO, as well as those which can be shifted by terms which vanish at the equation of motion (marked by an asterisk):
\begin{table*}[h]
\begin{center}
\begin{tabular}{c c c c c }
 \hline
 \hline
 i & $W_{i}^{(1)}$ & $a_{1\,i}^{(1)}$ & $a_{1\,i}^{(1)}(N=3)$ & EOM \\
 \hline
 5 & $\ltr \De \{ \cp, \usq\} \rtr$ & $-N/2$ & $-3/2$ & \\
 7 & $\ltr \De \cp \rtr \ltr \usq \rtr$ & $3/4 + N/8$ & $9/8$ & \\
 8 & $\ltr \De \usq \rtr \ltr \cp \rtr$ & $-1/4 + N/4$ & $1/2$ & \\
 9 & $\ltr \De [ \cm, \usq ] \rtr$ & $-N/4$  & $-3/4$ & $\ast$\\
 10 & $\ltr \De \cp^{2} \rtr$  & $-3/N + N/4$ & $-1/4$ & \\
 11 & $\ltr \De \cp \rtr \ltr \cp \rtr$ & $-1/2 -2/N^{2}$ & $-13/18$ & \\
 12 & $\ltr \De \cm^{2} \rtr$ & $0$ & $0$ & $\ast$ \\
 13 & $\ltr \De \cm \rtr \ltr \cm \rtr$ & $0$ & $0$ & $\ast$\\
 23 & $\ltr \Dem \{ \cm, \uum \} \rtr$ &  $0$ & $0$ & $\ast$\\
 36 & $\ltr \De[\cp,\cm] \rtr$ & $-1/N + N/4$ & $5/12$ & $\ast$ \\
 \hline
 \hline
\end{tabular}
\end{center}
\caption{List of operators needed for the \ktpp amplitude as well as those which can be 
shifted by EOM terms, marked by an asterisk. The $a_{1\,i}^{(1)}$ are given for Minkowski space-time. $N$ corresponds to the number of flavors.}
\label{t:wc}
\end{table*}
The operators given above differ slightly from the original EKW basis, which used
$ W_{36} = \ltr \De( [\cp,\cm] + \cp^{2} - \cm^{2} )\rtr$.
We use the above definition of $W_{36}$ since it simplifies the discussion about the operators which vanish at the solution of the equation of motion in section \ref{s:eomt} somewhat.
\subsection{NNLO: $Z^{(2)}_{\tiny{w}}$}
\begin{figure}[h]

\begin{center}

\begin{picture}(450,300)(0,0)

\CArc(50,250)(25,0,360)
\CArc(100,250)(25,0,360)
\CBox(65,240)(85,260){Black}{Yellow}
\Text(75,250)[]{$\mathfrak{w}^{4}_{0}$}
\Text(75,215)[]{\small{a}}

\CArc(200,300)(70.7,225,315)
\CArc(200,200)(70.7,45,135)
\Line(150,250)(250,250)
\CBox(140,240)(160,260){Black}{Yellow}
\CBox(240,240)(260,260){Black}{Yellow}
\Text(150,250)[]{$\mathfrak{w}^{3}_{0}$}
\Text(250,250)[]{$\mathfrak{s}^{3}_{0}$}
\Text(200,215)[]{\small{b}}

\CArc(300,250)(25,0,360)
\Line(325,250)(375,250)
\CArc(400,250)(25,0,360)
\CBox(315,240)(335,260){Black}{Yellow}
\Text(325,250)[]{$\mathfrak{w}^{3}_{0}$}
\CBox(365,240)(385,260){Black}{Yellow}
\Text(375,250)[]{$\mathfrak{s}^{3}_{0}$}
\Text(350,215)[]{\small{c}}

\CArc(75,175)(25,0,360)
\CBox(40,165)(60,185){Black}{Yellow}
\Text(50,175)[]{$\mathfrak{w}^{2}_{1}$}
\Text(75,140)[]{\small{d}}

\CArc(150,175)(25,0,360)
\CBox(115,165)(135,185){Black}{Yellow}
\Text(125,175)[]{$\mathfrak{w}^{2}_{0}$}
\Line(175,175)(225,175)
\CBox(165,165)(185,185){Black}{Yellow}
\Text(175,175)[]{$\mathfrak{s}^{3}_{0}$}
\CBox(215,165)(235,185){Black}{Yellow}
\Text(225,175)[]{$\mathfrak{s}^{1}_{1}$}
\Text(200,140)[]{\small{e}}

\CArc(300,175)(25,0,360)
\CBox(265,165)(285,185){Black}{Yellow}
\Text(275,175)[]{$\mathfrak{w}^{2}_{0}$}
\CArc(400,175)(25,0,360)
\Line(325,175)(375,175)
\CBox(315,165)(335,185){Black}{Yellow}
\Text(325,175)[]{$\mathfrak{s}^{3}_{0}$}
\CBox(365,165)(385,185){Black}{Yellow}
\Text(375,175)[]{$\mathfrak{s}^{3}_{0}$}
\Text(350,140)[]{\small{f}}

\CArc(75,100)(25,0,360)
\CBox(40,90)(60,110){Black}{Yellow}
\Text(50,100)[]{$\mathfrak{w}^{2}_{0}$}
\CBox(90,90)(110,110){Black}{Yellow}
\Text(100,100)[]{$\mathfrak{s}^{2}_{1}$}
\Text(75,65)[]{\small{g}}

\CArc(150,100)(25,0,360)
\Line(175,100)(225,100)
\CBox(165,90)(185,110){Black}{Yellow}
\CBox(215,90)(235,110){Black}{Yellow}
\Text(175,100)[]{$\mathfrak{w}^{3}_{0}$}
\Text(225,100)[]{$\mathfrak{s}^{1}_{1}$}
\Text(200,65)[]{\small{h}}

\CArc(400,100)(25,0,360)
\Line(325,100)(375,100)
\CBox(315,90)(335,110){Black}{Yellow}
\Text(325,100)[]{$\mathfrak{w}^{1}_{1}$}
\CBox(365,90)(385,110){Black}{Yellow}
\Text(375,100)[]{$\mathfrak{s}^{3}_{0}$}
\Text(350,65)[]{\small{i}}

\Line(112.5,25)(162.5,25)
\CBox(102.5,15)(122.5,35){Black}{Yellow}
\CBox(152.5,15)(172.5,35){Black}{Yellow}
\Text(112.5,25)[]{$\mathfrak{w}^{1}_{1}$}
\Text(162.5,25)[]{$\mathfrak{s}^{1}_{1}$}
\Text(137.5,-10)[]{\small{j}}

\CBox(262.5,15)(282.5,35){Black}{Yellow}
\Text(273.5,25)[]{$\bar{S}_{2}^{\tiny{w}}$}
\Text(272.5,-10)[]{\small{k}}

\end{picture}

\end{center}
\caption{Diagrams contributing to the generating functional $ Z^{(2)}_{\tiny{w}}$.
They split into the class of 1PI
diagrams (a,b,d,g), the 1PR diagrams (c,e,f,h,i,j), and the \ho 2
action $ \bar{S}_{2}^{\tiny{w}} $, diagram k.}
\label{f:tlc}
\end{figure}
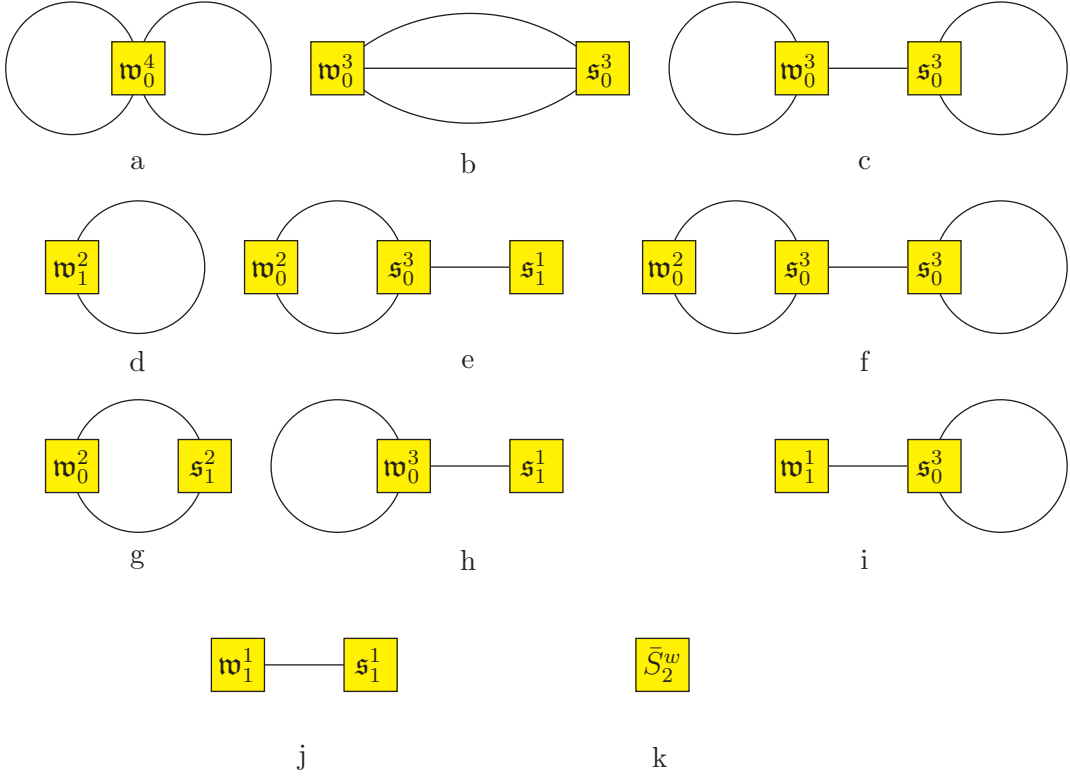

At \ho 2, we have the following diagrams:
\bea
  Z^{(2)}_{\tiny{w}}		     & = & -\frac{1}{6} \wvb_{0}^{ijk}G_{ir}G_{js}G_{kt}\svb_{0}^{rst}
                                   +\frac{1}{8} G_{ij}\wvb_{0}^{ijkl}G_{kl}
                                   +\frac{1}{2}  \wvb_{1}^{ij}G_{ij} 
				   -\frac{1}{2} \wvb^{ik}_{0}G_{ij}G_{kl}\svb^{jl}_{1} \nn
                             &   & -\frac{1}{4} G_{ij}\wvb_{0}^{ijk}G_{kr}\svb_{0}^{rst}G_{st}
 				   + \frac{1}{4} \wvb^{ij}_{0}G_{ik}G_{jl}\svb^{jkm}_{0}G_{mn}\svb^{mrs}_{0}G_{rs}
			     	   -\frac{1}{2}   \wvb_{1}^{i}G_{ir}\svb_{0}^{rst}G_{st}\nn
			     &  &  -\frac{1}{2} G_{ij}\wvb_{0}^{ijk}G_{kr}\svb_{1}^{r}
				   +\frac{1}{2} \wvb^{ik}_{0}G_{ij}G_{kl}\svb^{jlm}_{0}G_{mn}\svb^{n}_{1} 
			           - \wvb_{4}^{i}G_{ir}\svb_{1}^{r}
			           + \bar{S}^{\tiny{w}}_{2} \nn
			     &  &  + \mO(G_{F}^{2}) \label{eq:z2} \co
\eea
where $G_{ij}$ is the propagator corresponding to the $\xi$ field, whose ultraviolet divergent part is 
provided in appendix \ref{app:hkm}. The subscript of the vertices  denotes their \ho. The diagrams corresponding to Eq. (\ref{eq:z2}) are drawn in Fig. \ref{f:tlc}.

\noindent The NNLO Lagrangian $\mL^{(2)}_{\tiny{w}}$, represented by diagram k in
Fig. \ref{f:tlc},  has to cancel the divergences 
which are generated from the loop part of  $Z^{(2)}_{\tiny{w}}$. It takes the form: 
\bea
 \mL^{(2)}_{\tiny{w}} 
 & = & C g_{8} \sum_{i} c_{i}^{(2)} W^{(2)}_{i} \co
 \label{eq:wl2}
\eea
with the bare LEC's:
\be
 c_{i}^{(2)} =  (\mu c)^{-2\eps} \Big( c_{i}^{(2)r}\big(\mu,\eps\big) +  a^{(2)}_{1\,i}\big(\vec{c}^{(1)}(\mu,\eps)\big) \Lambda 
 +  a^{(2)}_{2\,i} \Lambda^{2}  \Big) \fs 
 \label{eq:wlec2}
\ee
\subsubsection{The connection between $\mA^{(2)}_{2}$ and the double chiral logs}
The highest pole of $\mL^{(2)}$, $\mA^{(2)}_{2} := \sum a^{(2)}_{2\,i} W^{(2)}_{i}$, can be used to calculate the double chiral logs which are generated from
genuine two loop diagrams of a process under consideration. Let us outline how this
works: The sum of all diagrams in Fig. \ref{f:tlc} with the exception of the counterterm diagram k will result in an expression which is proportional to the square of $\mQ$, Eq. (\ref{eq:qs}) plus other contributions which are not related to
double logs (abbreviated by the dots):
\be
 C g_{8} \sum_{i} \alpha_{i} W_{i}^{(2)} \mQ^{2} + ... = 
  C  g_{8} \sum_{i} \alpha_{i} W_{i}^{(2)} \mu^{-2\eps}\Big( \Lambda^{2} - \Lambda \hat{N} \log(\frac{m^{2}}{\mu^{2}}) + \big( \frac{\hat{N}}{2} \log(\frac{m^{2}}{\mu^{2}})\big)^{2} \Big) + ... \fs \no
\ee 
The $\Lambda^{2}$ divergences above have to be canceled by $\mA^{(2)}_{2}$, which translates into the following identity:
\bea
 C g_{8} \mu^{-2\eps} \sum_{i} \alpha_{i}W_{i}^{(2)} 
 \big( \frac{\hat{N}}{2} \log(\frac{m^{2}}{\mu^{2}})\big)^{2} & = & 
 - C g_{8} \mu^{-2\eps} \sum_{i} a^{(2)}_{2\,i} W_{i}^{(2)}
 \big( \frac{\hat{N}}{2} \log(\frac{m^{2}}{\mu^{2}})\big)^{2}  \con 
 & = & - C g_{8} \mu^{-2\eps} \mA^{(2)}_{2}
 \big( \frac{\hat{N}}{2} \log(\frac{m^{2}}{\mu^{2}})\big)^{2} \fs \no
\eea
In addition to the "genuine" double logs above, there are also contributions from 
one particle reducible topologies, the LSZ reduction, plus shifts of bare parameters
to their renormalized values in lower order contributions of the process under consideration.
\setcounter{equation}{0}
\section{Equation of motion terms}
\label{s:eomt}
In \cite{Buchler:2003vw} it was shown that one has to allow for operators which vanish at
the solution of the equation of motion (EOM terms in short)  
to define a basis in which one-particle-reducible (1PR) graphs contributing to the generating functional do not generate divergences.
The equation of motion in euclidean space-time reads:
\bea
 \xh & := & \Dm \um + \frac{\i}{2} \cmh  = 0 \co
 \label{eq:eom}
\eea
with: 
\bea
\Dm \um & := & \partial_{\mu} \um + [ \Gamma_{\mu},\um] \co \nn
\Gamma\dom & := & \frac{1}{2} \big( u^{\dagger}(\pa\dom - \i r\dom )                                  + u(\pa\dom - \i l\dom) u^{\dagger} \big)  \co \nn
\cmh & := & \cm - \ltr \cm \rtr / N \fs \no
\eea
Before we can start with the actual calculation discussed in 
section \ref{s:rge} and \ref{s:oc}, we need to define this proper basis, 
relevant for our computation:  \\
Defining the generating functional 
of proper vertices:
\be
 \bar{\Gamma}^{(n)}_{\tiny{a}}[J]  =  \Gamma^{(n)}_{\tiny{a}}[J,\phi]_{\phi = \bar{\phi}} 
 := Z^{(n)\tiny{\mbox{1PI}}}_{\tiny{a}}[J] \quad ; \; a = s,w  \con 
\ee
evaluated at the solution of the EOM, and its functional derivative
\be
 \bar{\Gamma}^{(n)\,i}_{\tiny{a}}[J]  =  ( \Gamma^{(n)\,i}_{\tiny{a}}[J,\phi] )_{\phi = \bar{\phi}} \quad ; \; a = s,w \no \co
\ee
we can decompose $Z^{(2)}_{\tiny{w}}$ into a 1PI and 1PR part: 
\be
Z_{\tiny{w}}^{(2)} = Z_{\tiny{w}}^{(2)\tiny{\mbox{1PI}}} + Z_{\tiny{w}}^{(2)\tiny{\mbox{1PR}}} = \bar{\Gamma}_{\tiny{w}}^{(2)}  -\bar{\Gamma}_{\tiny{w}}^{(1)\,i} G_{ij} \bar{\Gamma}_{\tiny{s}}^{(1)\,j} \fs
\label{eq:z2dec}
\ee
As Eq. (\ref{eq:z2dec}) illustrates, the 1PR portion of $Z$ will not contribute to divergences 
if in addition to $\Gamma$ itself all its functional derivatives  are finite.
The latter can be achieved by appropriate additions of EOM terms to the Lagrangian \cite{Buchler:2003vw} 
(We will however conclude this section with a stronger proposition on this point)
.\\
In the strong sector, we have at \ho 1:
\be 
 \hat{\mL}^{(1)}_{\tiny{s}} := \mL_{\tiny{\mbox{GL}}} + x_{1}^{(1)} \ltr \cm \xh \rtr
 + x_{2}^{(1)} \ltr \xh \xh \rtr \co 
\ee
with $\mL_{\tiny{\mbox{GL}}}$ the usual Gasser-Leutwyler Lagrangian 
\cite{Gasser:1984yg,Gasser:1985gg} and two additional EOM terms with coefficients $x_{1}^{(1)},x_{2}^{(1)}$.  \\
It was shown in \cite{Bijnens:1998yu} that the $\Gamma^{(1)\,i}_{\tiny{s}}$ is finite if one discards the EOM terms altogether (i.e. $x_{1}^{(1)}=x_{2}^{(1)}=0$), 
which is sufficient for our purposes. This result is to be expected since the only building block in $\mL_{\tiny{\mbox{GL}}}$ which corresponds to EOM terms, $\ltr \cm \cm \rtr$, has a vanishing divergent counterterm.  \\
In the weak sector, things get a little bit more involved. In addition to the EKW Lagrangian given in Eq. (\ref{eq:wcl1}), we have six EOM terms:
\be
 \hat{\mL}_{\tiny{w}} ^{(1)} = C g_{8}F^{2}_{0}\Big( \sum_{i=1}^{37} c_{i}^{(1)}W_{i}^{(1)} + \sum_{i = 1}^{6} e_{i}^{(1)} E_{i}^{(1)}\Big) \co
\ee
listed in Table \ref{t:eomt}: \\
\begin{table*}[h]
\begin{center}
\begin{tabular}{c c c}
\hline
\hline
i & $E_{i}^{(1)}$ & $k\;(W_{k}^{(1)})$\\
\hline
1 & $\i \ltr \De [ \usq, \xh ]\rtr$ & 9\\
2 & $\i \ltr \De \{ \cm, \xh \} \rtr$ & 12\\
3 & $ \ltr \De \xh \xh \rtr$ & 12\\
4 & $ \i \ltr \De \xh \rtr \ltr \cm \rtr$ & 13\\
5 & $\ltr \Dem \{  \um , \xh \} \rtr$ & 23\\
6 & $\i \ltr \De [ \cp, \xh ] \rtr$ & 36\\
\hline
\hline
\end{tabular}
\end{center}
\caption{All EOM terms for $\mL^{(1)}_{\tiny{w}}$.
In the last column we provide the operators of the EKW basis, given in section \ref{s:gfnwl}, Table \ref{t:wc},  
which corresponds to the respective EOM term.}
\label{t:eomt}
\end{table*}
\\
\noindent In order to pin down the coefficients $e_{i}^{(1)}$, we calculate the  functional derivative $\bar{\Gamma}^{(1)\,i}_{\tiny{w}} = Z^{(1)\,i}_{\tiny{w}}$ explicitly:
:
\bea
 \bar{\Gamma}^{(1) \, i }_{\tiny{w}} & = & \frac{1}{2} \wvb^{ijk}_{0} G_{jk}
 - \frac{1}{2} \svb^{ijk}_{0} G_{jl} G_{k m} \wvb^{lm}_{0}  +  \wvb^{i}_{1} \fs
\label{eq:fd}
\eea
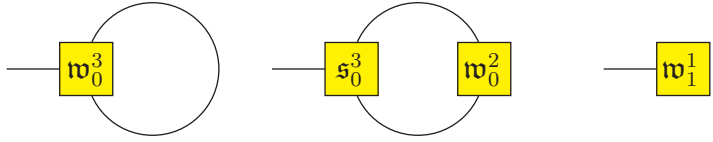
\begin{figure}[th]
\begin{center}
\begin{picture}(150,100)(100,25)

\Line(20,75)(40,75)
\CArc(75,75)(25,0,360)
\CBox(40,65)(60,85){Black}{Yellow}
\Text(50,75)[]{$\mathfrak{w}^{3}_{0}$}
\Line(120,75)(140,75)
\CArc(175,75)(25,0,360)
\CBox(140,65)(160,85){Black}{Yellow}
\Text(150,75)[]{$\mathfrak{s}^{3}_{0}$}
\CBox(190,65)(210,85){Black}{Yellow}
\Text(200,75)[]{$\mathfrak{w}^{2}_{0}$}

\Line(245,75)(265,75)
\CBox(265,65)(285,85){Black}{Yellow}
\Text(275,75)[]{$\mathfrak{w}^{1}_{1}$}

\end{picture}
\end{center}
\caption{Graphical representation of Eq. \ref{eq:fd}}
\label{f:fd}
\end{figure}
The diagrammatic representation of Eq. (\ref{eq:fd}) is shown in Fig. \ref{f:fd}.
To compute the ultraviolet divergent part of these diagrams, we have to expand the lowest order Lagrangians, Eqs. (\ref{eq:loscl}) and (\ref{eq:lowcl}), together, with the counterterm Lagrangian $\mA_{1}^{1} = \sum Z_{i}W_{i}$,
Eq. \ref{eq:wcl1}.
In appendix \ref{app:bbexp} we give the expansion of the building blocks defined in \ref{eq:nocl}
in terms of the quantum fluctuations $\xi$ and the background fields. \\
The Wick contractions 
are then performed using the heat kernel representation of propagators, 
briefly explained in appendix \ref{app:hkm}.
For the tadpole diagram $\wvb^{ijk}_{0} G_{jk}$, one uses the identities given in
Eq. (\ref{eq:hkpro}), whereas for the diagram $\svb^{ijk}_{0} G_{jl} G_{k m} \wvb^{lm}_{0}$, one employs the identities for products of propagators, provided in Eqs. (\ref{eq:expjo}),(\ref{eq:propfunc}))
and (\ref{eq:hkft}). A more explicit discussion of how the computation works is 
given in section \ref{s:oc}. \\
The calculation, whose result is too lengthy to be displayed here,  
 yields for the coefficients $e_{i}^{(1)}$ ( euclidean space-time, $N$ is the number of flavors):
\be
 e_{1}^{(1)} = \frac{N}{2} \; ; \;  e_{2}^{(1)} = e_{3}^{(1)} = 
 e_{4}^{(1)} = e_{5}^{(1)}  = 0 \; ; \; e_{6}^{(1)} = 
 - \frac{N}{2} + \frac{2}{N} \fs  
\ee
This shift of the original EKW Lagrangian corresponds to a replacement of the following building blocks:
\bea
 W_{9} = \ltr \De [ \cm, \usq ] \rtr & \longrightarrow &
  W_{9}' \cdot 2\i : =  \ltr \De [ \Dm \um , \usq ] \rtr \cdot 2\i \co \nn
 W_{36} = \ltr \De [\cp, \cm ] \rtr  & \longrightarrow &
  W_{36}' \cdot 2\i : = \ltr \De [ \cp , \Dm \um ] \rtr \cdot 2\i \fs
 \label{eq:bsh}
\eea
It is striking that $W_{9}$ and $W_{36}$ are the only operators which can be shifted by
EOM terms \textit{and} have a non-vanishing counterterm coefficient $a_{i}^{(1)}$ (see Table \ref{t:wc}). This observation leads to the conjecture that $\Dm \um$ instead of $\cmh$ should be used in loop calculations. A more general discussion of this point will be given in a separate 
paper \cite{eom}.  \\
\setcounter{equation}{0}
\section{The calculation of $\mA^{(2)}_{2}$}
\label{s:rge}
For the computation of $\mA^{(2)}_{2}$, we use renormalization group techniques:
If we write $\mL^{(2)}_{\tiny{w}}$, given in Eq. (\ref{eq:wl2}), split into the renormalized and
counterterm part as follows:
\be
 \mL^{(2)}_{\tiny{w}} =
C g_{8} \big( \mL^{(2)\;r}_{\tiny{w}} + 
\mA^{(2)}_{1} \Lambda + \mA^{(2)}_{2} \Lambda^{2} \big) \co \nn  
\label{eq:wl2b}
\ee
the RGE imply the identity: 
($ \partial^{(n)}_{i} = \partial / \partial c^{(n)r}_{i}$):
\bea
 \label{eq:gtl2}
  \mathcal{A}^{(2)}_{2} & = &  \frac{1}{2} \vec{a}^{(1)}_{1} \vec{\partial}^{(1)} \mathcal{A}^{(2)}_{1}   \co
\eea
with $\vec{a}^{(1)}_{1}$ being the counterterm of $\mL^{(1)}_{\tiny{w}}$ defined in Eq. (\ref{eq:wlec1}).
For the case of the weak nonleptonic sector, Fig. \ref{f:tlc} shows  all diagrams which can contribute to divergences at \ho 2.
If we act with the operator on the RHS of Eq. (\ref{eq:gtl2}), $\vec{a}^{(1)}_{1} \vec{\partial}^{(1)}$, on all these diagrams
and neglect the 1PR topologies (see section \ref{s:eomt}), there is only diagram d which can give a non-vanishing contribution.   
The general diagrammatic representation of this statement, Eq. (\ref{eq:gtl2}), 
is shown in Fig. \ref{fig:wdp}.
A thorough discussion and derivation of this RGE approach can be found in \cite{Buchler:2003vw}. The specific result, Eq. (\ref{eq:gtl2}), was already used
 in \cite{Weinberg:1979kz}. \\
Due to the RGE, we can therefore obtain the leading poles at \ho 2 by computing only the one loop diagram d in Fig. \ref{f:tlc} and weighting it with the factor $1/2$, 
instead of having to compute the genuine 1PI two loop diagrams a and b. 
 \begin{figure}[h]
\begin{eqnarray*}
\nonumber
 \quad \;
 \begin{picture}(40,40)(0,-2)
 \CBox(-11,-11)(12,11){Black}{Yellow}
 \Text(1,0)[]{$\mA^{(2)}_{2}$}
\end{picture}
& = & \frac{1}{2}
 \begin{picture}(40,40)(-40,-2)
 \CArc(25,0)(25,0,360)
 \CBox(-11,-11)(12,11){Black}{Yellow}
 \Text(1,0)[]{$\mA^{(1)}_{1}$}
\end{picture}
\end{eqnarray*}
\vskip 0.5cm
\caption{Calculation of $ \mA^{(2)}_{2} $ with the RGE.}
\label{fig:wdp}
\end{figure}

\noindent In appendix \ref{s:a22} we provide all the operators of $\mA_{2}^{(2)}$
 which can contribute to 
amplitudes involving at most four pseudo-Goldstone particles and no vector-sources, denoted by 
$\tilde{\mA}^{(2)}_{2}$. This result has been used for the calculation of the double log contribution to the \ktpp amplitude, which is presented in a separate paper \cite{dlkspp}. 
The full expression is approximately four times the size of the truncated one shown in the appendix,  and thus too lengthy to be displayed in this paper. It can however be obtained from the author. An overview about the workings of the actual computation can be found in section \ref{s:oc}.
\setcounter{equation}{0}
\section{Outline of the computation}
\label{s:oc}
In this section we provide a sketch of how the whole calculational machinery
used in this paper works.
\subsection{Tadpole graphs}
We will illustrate the computation of a tadpole graph by computing a part of $\mA^{(2)}_{2}$ with the help of Eq. (\ref{eq:gtl2}).
Our starting point is the counterterm of the NLO Lagrangian $\mL_{\tiny{w}}^{(1)}$ \cite{Kambor:1990tz,Ecker:1993de}.
For completeness we provide the expanded building blocks needed in appendix \ref{app:bbexp}.
We decided, however, not to reproduce the whole expanded expression of $\mA^{(1)}_{1}$ here, since it is rather long. \\
We will rather restrict ourselves to the first building block which occurs in $\mA^{(1)}_{1}$,
$W_{1} = \ltr \De \usq \usq \rtr$:
We expand $W_{1}$ in the quantum fluctuation fields $\xi$:
\bea
 \ltr \De \usq \usq \rtr & = & \ltr \Deb \usqb \usqb \rtr \nnn
 + \frac{\i}{2} \ltr \Deb [ \usqb \usqb , \lai] \rtr \xi^{i}
 - \ltr \Deb \{\usqb,\{\umb,\lai \}\} \rtr \xi_{\mu}^{i} \nnn
 -\frac{1}{8} \ltr \Deb \big( \{\usqb,\{\umb,[[\umb,\lai],\laj]\}\}
  + [[\usqb\usqb,\lai],\laj]  \big) \rtr \xi^{i} \xi^{j}
 \nnn
 -\frac{\i}{2} \ltr \Deb [\{\usqb,\{\umb,\lai\}\},\laj] \rtr \xi_{\mu}^{i} \xi^{j} \nnn +
 \ltr \Deb \big( \{ \usqb, \lai \laj \} + \{\umb , \lai\}\{ \unb,\laj \}\big)
  \rtr \xi_{\mu}^{i} \xi_{\nu}^{j} + \mO(\xi^{3}) \fs
 \label{eq:w1exp}
\eea
Due to delta functions generated by functional differentiation, no space-time
integration survives and all building blocks and $\xi$-fields in Eq. (\ref{eq:w1exp}) are
evaluated at the same space-time point $x$. Indices in $SU(N)$ space are denoted by
$i,j$.
The $\xi$-fields of the bilinear terms proportional to $\xi^{i}\xi^{j}$, $\xi^{i} \xi_{\mu}^{j}$ and
$\xi_{\mu}^{i} \xi_{\nu}^{j}$ are then contracted with the heat kernel representations
 of $G_{\Delta}(x,x)$, $d_\mu^x G_\Delta(x,y)_{|_{y=x}}$ and
$d_\mu^x d_\nu^y G_\Delta(x,y)_{|_{y=x}}$ respectively, listed in Eq.
(\ref{eq:hkpro}). For the first contributing term in Eq. (\ref{eq:w1exp}) we get the following contribution to the action $\bar{S}_{2}^{\tiny{w}}$:
\bea
& &  -\frac{1}{8} \int d^{d}x \ltr \Deb \big( \{\usqb,\{\umb,[[\umb,\lai],\laj]\}\}
  + [[\usqb\usqb,\lai],\laj] \big) \rtr G_{\Delta}(x,x)_{ij} \nn
& = & -  \frac{(c\mu)^{-\eps}\Lambda}{4}
     \int d^{d}x \ltr \Deb \big( \{\usqb,\{\umb,[[\umb,\lai],\laj]\}\}
  + [[\usqb\usqb,\lai],\laj] \big) \rtr  (a_{1}^{\De})_{ij} + \mbox{finite terms} \co \nn
& = &  -  \frac{(c\mu)^{-\eps}\Lambda}{4}
    \int d^{d}x \ltr \Deb \big( \{\usqb,\{\umb,[[\umb,\lai],\laj]\}\}
  + [[\usqb\usqb,\lai],\laj] \big) \rtr  \cdot \nn &   &
 - \frac{1}{8} \ltr [\um,\lambda_{i}][\um,\lambda_{j}] + \{\lambda_{i},\lambda_{j}\}\cp \rtr + \mbox{f.t.} \fs
\eea
For the subsequent contraction of the $SU(N)$ indices $i,j$ one uses the completeness relations:
\bea
 \sum_{i=1}^{N^{2}-1}\ltr \lambda_{i} A \lai B \rtr = - \frac{2}{N}\ltr A B \rtr + 2 \ltr A \rtr \ltr B \rtr \co \nn
 \sum_{i=1}^{N^{2}-1} \ltr \lambda_{i} A \rtr \ltr \lai B \rtr = 2 \ltr A B \rtr
 - \frac{2}{N} \ltr A \rtr \ltr B \rtr \fs
 \label{eq:cr}
\eea
This last step of the computation is obviously straightforward, and we forbear to
display the final result, since it is again rather lengthy.

\subsection{Beyond the tadpole}
The computation of the functional derivative of $\Gamma_{\tiny{w}}^{(1)}$, Eq. (\ref{eq:fd}),
involves a diagram with two propagators and is therefore a little bit more involved: in addition to
the $SU(N)$ contractions one has to deal with the space-time dependent part of the product of the two propagators. Let us again
restrict ourselves to the simple case of a part of the computation where the vertices do not carry any derivatives acting on the  $\xi$-fields. The structure of such a piece is then:
\be
 Q^{a} = \int d^{d}x d^{d}y \, v^{ajk}_{\tiny{s}}(x) G_{jl}(x,y)G_{km}(x,y) v^{lm}_{\tiny{w}}(y)  \fs
\ee
Here $i,j,k,l,m$ are again pure $SU(N)$ indices, and $a$ corresponds to the space-time point $x_{i}$ as well as to the $SU(N)$ index $i$.
The space-time dependent part can be evaluated with the help of Eqs. (\ref{eq:propfunc}) and
(\ref{eq:hkft}) (We suppress the vertices $v^{ajk}_{\tiny{s}}(x)$ and $v^{lm}_{\tiny{w}}(y)$ in this step):
\bea
\int d^{d}x d^{d}y  G_{jl}(x,y)G_{km}(x,y) & = &
\Big( 4 \hat{N} \frac{\Gamma(1 - \eps/2)}{\pi^{-\eps/2}} \Big)^{2} \cdot \nn
& & \int d^{d}x d^{d}y\, a_{0}^{\De}(x,y)_{jl} a_{0}^{\De}(x,y)_{km} |x-y|^{-d + \eps } + \mbox{f.t.} \co \nn
& = & \Big( 4 \hat{N} \frac{\Gamma(1 - \eps/2)}{\pi^{-\eps/2}} \Big)^{2}
(\pi)^{d/2} \frac{\Gamma(\eps/2)}{\Gamma(d/2)} \cdot \nn
& &  \int d^{d}x d^{d}y \, a_{0}^{\De}(x,y)_{jl} a_{0}^{\De}(x,y)_{km}
 \mu^{-\eps}\delta^{d}(x-y) + \mbox{f.t.} \co \nn
& = & 2 (c \mu )^{-\eps} \Lambda  \int d^{d}x \, \delta_{jl} \delta_{km} + \mbox{f.t.} \fs
 \label{eq:mp}
\eea
Reinserting the vertices again, we finally get ($v^{a j k }(x) = \delta^{d}(x-x_{i}) v^{ijk}$(x)):
\bea
 Q^{a} = Q^{i}(x_{i}) & = &
  2 (c \mu )^{-\eps} \Lambda \, \int d^{d}x  \,v^{ajk}_{\tiny{s}}(x)v^{jk}_{\tiny{w}}(x) + \mbox{f.t.} \con
  & = & 2 (c \mu )^{-\eps} \Lambda \,  \,v^{ijk}_{\tiny{s}}(x_{i})v^{jk}_{\tiny{w}}(x_{i}) + \mbox{f.t.} \fs
\eea
If $\xi$-fields with derivatives are contracted, one will in addition generate
derivatives acting on the delta-function in Eq. (\ref{eq:mp}).
These derivatives can
be shifted to the vertices and Seeley-DeWitt coefficients by partial integration.
As a last step, we will again have
to contract the $SU(N)$ indices of the vertices with the one's of Seeley-DeWitt
coefficients of the expansion of the propagator, Eq. (\ref{eq:expjo}), using once more the 
completeness relations in Eq. (\ref{eq:cr}).
\subsection{Verification of the calculation}
This computation was exclusively performed with FORM 3.1 \cite{Vermaseren:2000nd}, a symbolic 
manipulation program. Since the whole calculation is thus fully automatized, one can 
conveniently adapt the code to problems whose solutions are known:  
In order to check the written code, we used it to recalculate two known counterterm 
Lagrangians.  We replaced the original Lagrangian $\mA^{(1)}_{1}$ (see Fig. \ref{fig:wdp}) with the respective Lagrangians required for their calculation 
as starting point,
 but left the rest of the code unchanged:
\begin{enumerate}
\item 
We recalculated the counterterms for $\mL_{\tiny{w}}^{(1)}$ using the method 
outlined in this paper, i.e. by the computation of the ring diagram 
$\frac{1}{2}\wvb^{ij}G_{ij}$ instead of using the logarithm:  
$\frac{1}{2} \mTr(\mln(\De_{\tiny{s}} +\De_{\tiny{w}}))$ and projecting out the part linear
in $G_{F}$, as employed in the original calculation \cite{Kambor:1990tz}. 
We found total agreement. 
\item
We recalculated the leading poles at \ho 2 in the strong sector and compared our result 
with the one obtained by Bijnens, Colangelo and Ecker \cite{Bijnens:1998yu,Bijnens:1999hw}. 
The outcome of our computation matched completely with their result.
\end{enumerate}
Since this is the very first NNLO calculation of \cpt in the weak sector, 
it was not possible to compare it directly with genuine two loop calculations. However, in our opinion, the two checks listed above, in particular the second one, are highly nontrivial, and yield sufficient evidence that the FORM code written for the computation is correct.
\section{Conclusions}
We have determined the leading divergences for the weak nonleptonic chiral Lagrangian at NNLO for the part which
transforms like an octet under the chiral group, extending the analogue computation of the NLO counterterms \cite{Kambor:1990tz,Ecker:1993de}.
The obtained result can be used to calculate double log contributions of observables, which at two loop order are the only quantities that do not depend on any LEC's \footnote{besides subtracted loop integrals which would require a fully-fledged two loop calculation.}.
Unlike in the strong sector, in the weak sector it is extremely difficult
to determine these LEC's, and presumably one will not be able to pin them down in the near future, if ever.
Thus, the double logs provide a first estimate about the NNLO corrections one has to expect,
without the need of these unknown LEC's as input. Typically, the double log contributions in three flavor \cpt
amount to around $10\%$ of the corrections to the lowest order result. \\
Corrections to lowest order \cpt quantities are used for chiral extrapolations of
lattice data. In these days, lattice simulations have entered a stage where one uses fully unquenched quarks, and aim to predict observables with an accuracy in the range of some percent. In view of such high
precision, it certainly makes sense to include NNLO corrections in these extrapolations.
The main objective of the present calculation
is the use of the counterterm Lagrangian for the computation of the corresponding double log contributions to the \ktpp amplitude in the $\De I = 1/2 $ channel, which is presented in another paper \cite{dlkspp}.
Other interesting applications are analogue calculations for the \ktppp and $K \to \pi \gamma \gamma$ amplitudes.
\section*{Acknowledgments}
I would like thank Gilberto Colangelo for participation in the early stages of this work, many very helpful discussions, and continuous encouragement. \\
This work was supported by the Swiss National Science Foundation. 
\appendix
\renewcommand{\theequation}{\thesection.\arabic{equation}}
\setcounter{equation}{0}
\section{Background field method}
\label{app:bfm}
In this appendix we provide a very brief outline of the background field formalism \cite{Abbott:1982ke}, 
which was used in the present calculation. 
The field $\vph$ in $U = \mexp(\i \sqrt{2} \phi/F)$ is split into a background
part $\bar{\phi} $, which is normally taken to be at the solution of the equation of motion, and a quantum fluctuation field $\xi$:
\bea
 \phi & = & \bar{\phi} + \xi/\sqrt{2} \fs
\eea
$\bar{\phi}$ will be used as an external field, i.e. it will not propagate; therefore, only $\xi$ can generate $\hbar$ corrections. With this substitution, 
the action assumes the form: 
\bea
 \label{eq:sexp}
 S[\phi] & \to & S[\bar{\phi}] +
                \frac{1}{2!}S[\bar{\phi}]^{ij}\xi_{i}\xi_{j}+
                \frac{1}{3!}S[\bar{\phi}]^{ijk}\xi_{i}\xi_{j}\xi_{k}
                + \frac{1}{4!}S[\bar{\phi}]^{ijkl}\xi_{i} \xi_{j}\xi_{k}\xi_{l} + ...  \fs
\eea
In the equation above we have assumed that $\bar{\phi}$ is a solution of the equation of motion,
so that the linear term in the expansion vanishes. The RHS of (\ref{eq:sexp}) can now be viewed
as the new Lagrangian where the new integration variables of the path integral are $\xi$ instead
of $\vph$.
The inverse of the bilinear operator in $\xi$, $S[\bar{\phi}]^{ij}(x,y)  = \delta^{d}(x - y)\De_{ij}$ corresponds
as usual to the propagator. Please note that we will only use the bilinear part of the strong chiral Lagrangian to define the propagator; the expanded form of the weak  nonleptonic chiral Lagrangian will only be needed for the definition of vertices, each insertion thereof 
corresponding to a factor $G_{\tiny{\mbox{F}}}$.
$\Delta$ is brought into the canonical form of an elliptical operator:
\bea
 \label{eq:cfop}
 \De_{ij} & = & (-d^{2}_{x} + \sigma(x))_{ij}   \co \\
 d_{\mu \, kl}  & = & \delta_{kl}\ddm + \gamma\dom(x)_{kl}   \co
\eea
which have the following explicit form for the \cpt Lagrangian, Eq. (\ref{eq:loscl}):
\bea
 \gamma_{\mu\,ij} & = & -\frac{1}{2} \ltr \Gamma_{\mu}[\lambda_{i},\lambda_{j}] \rtr \co \nn
 \sigma_{ij} & = & \frac{1}{8} \ltr [\um,\lambda_{i}][\um,\lambda_{j}] + \{\lambda_{i},\lambda_{j}\}\cp \rtr \fs
 \label{eq:sigma}
\eea
and will be used to define the propagator in appendix \ref{app:hkm}. The field strength 
associated to $d_{\mu}$: 
\be
 \gamma_{\mu\nu} = [d\dom,d\don] = \ddm\gamma_{\nu}-\ddn\gamma_{\mu}
+ [\gamma_{\mu},\gamma_{\nu}]   \co
\label{eq:gmn}
\ee
will also be needed, and takes the explicit form:
\begin{alignat}{2}
  \gamma_{\mu\nu\;ij} & =  -\frac{1}{2} \ltr \Gamma_{\mu\nu}[\lambda_{i},\lambda_{j}] \rtr \quad ;  \quad & \Gamma_{\mu\nu} & = \frac{1}{4} [\um,\un] - \frac{\i}{2} \fpdmn \fs
  \label{eq:gmne}
\end{alignat}
\setcounter{equation}{0}
\section{The heat-kernel method}
\label{app:hkm}
In this appendix we give a brief summary of the heat-kernel method as developed 
by Jack and Osborn \cite{Jack:1982hf} and provide 
a compilation of the results which are needed for the calculation. Throughout this 
section we work in  euclidean space-time. The presentation is to a large extent based 
on \cite{Jack:1982hf,Bijnens:1999hw} and some results provided in \cite{Ball:1989xg}. \\ 
Let us consider the propagator $G_{\De}(x,y)$ associated to $\De$, which in euclidean space-time is defined by:
\be
\label{eq:bcgf}
\De_{x}G_{\De}(x,y) = \delta(x-y) \fs
\ee
In the Schwinger representation, $G_{\De}(x,y)$ is written as an integral over the eigentime $\tau$:
 \bea
  \label{eq:proint}
  G_{\De}(x,y) & = & \int_{0}^{\infty}d\tau \rho(\tau,\eps) \mG_{\De}(x,y;\tau)  \fs
\eea
The kernel of the integral, $\mG_{\De}(x,y;\tau)$, satisfies the diffusion equation:
\be
\partial_{\tau} \mG_{\De}(x,y;\tau) = -\De_{x} \mG_{\De}(x,y;\tau)  \co
\ee
and the boundary condition $\mG_{\De}(x,y;0) = \delta(x-y)$. \\
$\mG_{\De}(x,y;\tau)$ can be expanded in the Seeley-DeWitt coefficients:
\be
\label{eq:hkae}
  \mG_{\De}(x,y;\tau)
   = \frac{1}{(4\pi\tau)^{\frac{d}{2}}}
 e^{\frac{|x-y|^{2}}{4\tau}}\sum_{j=0}^{\infty}a_{j}^{\De}(x,y)\tau^{j} \co
\ee
 and after solving the
integral (\ref{eq:proint}) with the regulator function $\rho(\eps,\tau) =
(4\pi\tau)^{-\frac{\eps}{2}}$ corresponding to dimensional regularization, one gets the asymptotic
formula:
\bea
G_\Delta(x,y) &=& G_0(x-y) a_0^\Delta(x,y)+ R_1(x-y;c\mu) a_1^\Delta(x,y)
 \nn
& & + R_2(x-y;c\mu) a_2^\Delta(x,y) + R_3(x-y;c\mu) a_3^\Delta(x,y) \nn
& & +                 \overline{G}_\Delta(x,y;c\mu )  \co
\label{eq:expjo}
\eea
with the coefficients:
\bea
G_0(x)&=& \hat{N} \frac{\Gamma(1-\frac{\eps}{2})}{4\pi^{-\frac{\eps}{2}}}|x|^{\eps-2}  \co \nn
R_1(x;c\mu)&=& 2 (c\mu)^{-\eps} \Lambda  +
   \hat{N} \frac{\Gamma (-\frac{\eps}{2})}{\pi^{\frac{\eps}{2}}} |x|^{\eps}  \co \nn
R_2(x;c\mu)&=&  \frac{|x|^2}{4} \Big( -2(c\mu)^{-\eps} \Lambda+
   \hat{N} \frac{\Gamma (-1-\frac{\eps}{2})}{\pi^{\frac{\eps}{2}}} |x|^{\eps} \Big)  \co \nn
R_3(x;c\mu)&=&  |x|^4 \Big( (c\mu)^{-\eps} \Lambda +
  \hat{N} \frac{ \Gamma (-2-\frac{\eps}{2})}{\pi^{\frac{\eps}{2}}} |x|^{\eps} \Big) \co
\label{eq:propfunc}
\eea
where $c$ parametrizes the renormalization prescription (
$\mln(c) =  \exp(-( \mln(4\pi) + \Gamma^{'}(1) + 1 )/2$ for $\overline{\mbox{MS}}$.).). \\
Eq. (\ref{eq:expjo}) is only valid  asymptotically for $\tau \to 0$ and can therefore only be used to
extract the ultraviolet behavior of the propagator.
The Seeley-DeWitt coefficients are given by:
\bea
a_{0}^{\De} & = & 1 \co \nn
a_{1}^{\De} & = & -\sigma \co \nn
a_{2}^{\De} & = & \frac{1}{12}\gamma\domn \gamma\domn +\frac{1}{2} \sigma^{2}
-\frac{1}{6}d\dom d\dom \sigma  \co
\eea
with $\sigma$ and the field strength $ \gamma_{\mu\nu} $ defined in Eq. (\ref{eq:sigma})
 and (\ref{eq:gmn}) respectively. \\
In order to calculate the divergences of the tadpole graphs like Fig. \ref{fig:wdp}, we need the
propagator with up to three derivatives, which can be calculated by use of Eq. (\ref{eq:propfunc}),
projecting out the space-time independent part:
\bea
G_{\Delta}(x,x) & = &
(c\mu)^{-\eps}\Lambda \, 2 a_1^{\Delta} (x,x) + \overline{G}_{\Delta} (x,x;c\mu) \co \nn
d_\mu^x G_\Delta(x,y)_{|_{y=x}} & = &
(c\mu)^{-\eps}\Lambda\, 2 d^x_\mu a_1^\Delta (x,y)_{|_{y=x}} +
d^x_\mu \overline{G}_\Delta (x,y;c\mu)_{|_{y=x}}     \co \nn
d_\mu^x d_\nu^y G_\Delta(x,y)_{|_{y=x}} & = &
(c\mu)^{-\eps}\Lambda \Big( 2 d^x_\mu d^y_\nu  a_1^\Delta (x,y)_{|_{y=x}} +
 \delta_{\mu \nu} a_2^\Delta (x,x) \Big) \nn
& & + d^x_\mu d^y_\nu \overline{G}_\Delta (x,y;c\mu)_{|_{y=x}} \co \nn
d_\mu^x d_\nu^x G_\Delta(x,y)_{|_{y=x}} & = &
(c\mu)^{-\eps}\Lambda \Big( 2 d^x_\mu d^x_\nu  a_1^\Delta (x,y)_{|_{y=x}} -
 \delta_{\mu \nu} a_2^\Delta (x,x) \Big) \nn
& & + d^x_\mu d^x_\nu \overline{G}_\Delta (x,y;c\mu)_{|_{y=x}} \co \nn
d_\mu^x d_\nu^x d_\rho^y G_\Delta(x,y)_{|_{y=x}} & = & (c\mu)^{-\eps}\Lambda \label{eq:hkpro} \\
& & \Big( 2 d_\mu^x d_\nu^x d_\rho^y a_1^\Delta (x,x)   \nn
& & - \big( \delta_{\mu\nu} d_\rho^y a_2^\Delta (x,x)-\delta_{\mu\rho}d_\nu^x a_2^\Delta (x,x)
      -\delta_{\nu\rho}d_\mu^x a_2^\Delta (x,x)\,\big)\,\Big)  \nn
&  & +
d_\mu^x d_\nu^x d_\rho^y \overline{G}_\Delta (x,y;c\mu)_{|_{y=x}} \fs\nonumber
\eea
For the calculation of the divergences of the functional derivative of of $Z^{(1)}_{\tiny{w}}$, Eq. (\ref{eq:fd}) or Fig. \ref{f:fd}, one needs to deal with products of propagators.
After a couple of manipulations, the space-time dependent part of such 
products can be brought into the form of a sum of terms proportional to:
\be
 \prod_{j=1}^{n} \partial_{\alpha_{j}}^{z} \frac{1}{|z|^{2m}} \quad ; \quad
 z := y-x \quad ; \quad \partial_{\mu}^{x} = - \partial_{\mu}^{z} \quad ; \quad 
 \partial_{\mu}^{y} =  \partial_{\mu}^{z} \fs \no 
\ee
Such terms can be represented by delta
functions via their Fourier transforms in $d$-space-time-dimension:
\be
 \label{eq:hkft}
 \int d^{d}z \frac{1}{|z|^{2m}}  =  \pi^{\frac{d}{2}} \frac{\Gamma(-( m - d/2))}
 {\Gamma(m)} \big( q^{2} \big)^{m - d/2}  \co
\ee
substituting $q^{2n - k\eps} \to \mu^{-k\eps}(- \partial^2)^{n}\delta^{d}(z)\; ; \; n,k \in \mathbb{N}_{0}$, from which the ultraviolet divergent parts can easily be extracted.
\setcounter{equation}{0}
\section{Explicit result for $\mA^{(2)}_{2}$}
\label{s:a22}
\bea
\tilde{\mA}^{2}_{2}  & = &
       + \ltr \De \cp \cp \cp \rtr  (  - \frac{21}{8} + \frac{6}{N^{2}} - \frac{N}{8} + \frac{5N^{2}}{32} )        
       + \ltr \De \cp \cp \rtr\ltr \cp \rtr  ( \frac{1}{2} - \frac{5}{N^{3}} - \frac{1}{N} + \frac{N}{2} - \frac{N^{2}}{32} ) \nnn
       + \ltr \De \cp \rtr\ltr \cp \cp \rtr  ( \frac{1}{4} - \frac{2}{N^{3}} + \frac{3}{8N} + \frac{5N}{32} - \frac{N^{2}}{32} )
       + \ltr \De \cm \rtr\ltr \cp \cm \rtr  (  - \frac{2}{N^{3}} - \frac{5}{4N} + \frac{N}{16} ) \nnn
       + \ltr \De \cm \rtr\ltr \cp \rtr\ltr \cm \rtr  ( \frac{2}{N^{4}} + \frac{1}{2N^{2}} ) 
 + \ltr \De \cp \rtr\ltr \cp \rtr\ltr \cp \rtr  ( \frac{1}{8} + \frac{2}{N^{4}} + \frac{9N^{2}}{8} - \frac{7}{16N} 
-      \frac{3N}{32} )
\nnn
       + \ltr \De \cm \cm \rtr\ltr \cp \rtr  (  - \frac{3}{N^{3}} - \frac{3}{8N} + \frac{3N}{32} )
       + \ltr \De \cm \cp \cm \rtr  (  - \frac{3}{8} + \frac{4}{N^{2}} + \frac{N^{2}}{32} ) \nnn      
       + \ltr \De \cp \rtr\ltr \cm \cm \rtr  (  - \frac{1}{8N} - \frac{N}{32} )
       + \ltr\{\De \cm \cm \cp\}\rtr  (  - \frac{1}{8} + \frac{2}{N^{2}} + \frac{N^{2}}{32} ) \nnn
       + \ltr\{\De \cm \cp\}\rtr\ltr \cm \rtr  (  - \frac{1}{N^3} - \frac{5}{8N} + \frac{N}{32} ) 
       + \ltr [ \De \cm \cp ]\rtr\ltr \cp \rtr  ( \frac{1}{2N^{3}} + \frac{5}{8N} - \frac{N}{16} ) \nnn 
       + \ltr  [\De \cm \cp \cp]\rtr  (  - \frac{1}{8} - \frac{1}{2N^{2}} )
      + \ltr \De \cpl{\mu}{\mu} \rtr\ltr \cp \rtr  ( \frac{3}{8} - \frac{1}{2N} + \frac{N}{16} - \frac{N^{2}}{16} ) \nnn
      + \ltr  [\De \cm \cpl{\mu}{\mu}]\rtr  (  - \frac{1}{4} + \frac{N^{2}}{16} )
        + \ltr \De u_{\mu} \cp u^{\mu} \rtr\ltr \cp \rtr  (  - \frac{9}{16} + \frac{7}{8N} - \frac{3N}{32} + \frac{3N^{2}}{32} )        
       \nnn
       + \ltr \De \cp \rtr\ltr \cpl{\mu}{\mu} \rtr  ( \frac{1}{16} + \frac{7}{8N} - \frac{N}{8} - \frac{N^{2}}{16} )
      + \ltr \De u_{\mu} \rtr\ltr \cp u^{\mu} \rtr\ltr \cp \rtr  ( \frac{1}{8} + \frac{5}{4N^{2}} + \frac{1}{8N} + \frac{N}{8}) \nnn
       + \ltr \De u_{\mu} \rtr\ltr \cp \cp u^{\mu} \rtr  (  - \frac{1}{4} - \frac{2}{N^{1}} + \frac{N}{32} + \frac{N^{2}}{16} )
       + \ltr \{  \De \cpl{\mu}{\mu} \cp \} \rtr  (  - \frac{3}{8} - \frac{N}{16} + \frac{N^{2}}{8} ) \nnn
      + \ltr \De \cp \rtr\ltr \cp u^{2} \rtr  (  - \frac{1}{16} - \frac{15}{8N} + \frac{3N}{4} + \frac{3N^{2}}{32} )
       + \ltr \De u_{\mu} \cp \cp u^{\mu} \rtr  (  - \frac{1}{4} + \frac{N}{16} ) \nnn
       + \ltr \De u^{2} \rtr\ltr \cp \rtr\ltr \cp \rtr  (  - \frac{3}{32} - \frac{7}{8N^{2}} + \frac{1}{8N} + \frac{3N}{32} )
       + \ltr \De u^{2} \rtr\ltr \cp \cp \rtr  (  - \frac{1}{4} + \frac{9}{8N} - \frac{N}{16} + \frac{N^{2}}{16} )\nnn
       + \ltr \De u^{2} \cp \rtr\ltr \cp \rtr  (  - \frac{3}{32} + \frac{1}{8N} - \frac{N}{64} + \frac{N^{2}}{64} ) 
       + \ltr \De \cp u^{2} \rtr\ltr \cp \rtr  (  - \frac{3}{32} + \frac{1}{8N} - \frac{N}{64} + \frac{N^{2}}{64} )\nnn
       + \ltr \De \cp u^{2} \cp \rtr  (  - \frac{1}{2}  - \frac{3 N^{2}}{32} )
       + \ltr \De \cp \cp \rtr\ltr u^{2} \rtr  (  - \frac{1}{8} - \frac{19}{8N} + \frac{17N}{32} + \frac{N^{2}}{32} ) \nnn
             + \ltr \De \cp \rtr\ltr \cp \rtr\ltr u^{2} \rtr  ( \frac{11}{32} + \frac{17}{8N^{2}} + \frac{1}{16N} - \frac{3N}{32} )
        + \ltr\{\De u_{\mu} \cp\}\rtr\ltr \cp u^{\mu} \rtr  ( \frac{3}{4N} - \frac{3N}{32} + \frac{N^{2}}{32} ) \nnn
	+ \ltr\{\De u^{2} \cp \cp\}\rtr  (  - \frac{7}{16} - \frac{N}{16} - \frac{5N^{2}}{64} )
        + \ltr\{\De u_{\mu} \cp u^{\mu} \cp\}\rtr  ( \frac{1}{2} + \frac{3N}{32} - \frac{3N^{2}}{16} ) \nnn
       + \ltr\{\De u^{2} \cp\}\rtr\ltr \cp \rtr  ( \frac{3}{16} + \frac{1}{16N} - \frac{N}{64} + \frac{N^{2}}{64} )
	       + \i\ltr  [\De \cpl{\mu}{} u^{\mu} \cp]\rtr  (  - \frac{9}{16} - \frac{N}{32} + \frac{N^{2}}{8} ) \nnn
       + \i\ltr  [\De u_{\mu} \cpl{}{\mu}]\rtr\ltr \cp \rtr  (  - \frac{3}{16} + \frac{1}{4N}
           - \frac{N}{8} +  \frac{N^{2}}{32} )
       + \i\ltr  [\De u_{\mu} \cpl{}{\mu} \cp]\rtr  (  - \frac{1}{4} + \frac{N^{2}}{32} ) \nnn
       + \i\ltr  [\De u_{\mu} \cp]\rtr\ltr \cpl{}{\mu} \rtr  (  - \frac{3}{4N} + \frac{5N}{32} 
       + \i\ltr  [\De u_{\mu} \cp \cpl{}{\mu}]\rtr  ( \frac{5}{16} + \frac{N}{32} )
	\no
\eea
\bea 
& & 
       + \ltr \De h_{\mu\nu}  h^{\mu\nu}  \rtr\ltr \cp \rtr  ( \frac{N}{32} - \frac{N^{2}}{32} )
       + \ltr \De h_{\mu\nu}  \cp h^{\mu\nu}  \rtr  (  - \frac{1}{8} + \frac{N}{16} ) 
       + \ltr \De \cp \rtr\ltr h_{\mu\nu}  h^{\mu\nu}  \rtr  (  - \frac{N}{16} - \frac{N^{2}}{32} ) \nnn
       + \ltr\{\De \cp h_{\mu\nu}  h^{\mu\nu} \}\rtr  ( \frac{1}{16} - \frac{N}{32} + \frac{N^{2}}{16} 
       + \ltr  [\De \cm h_{\mu\nu}  h^{\mu\nu} ]\rtr  ( \frac{N^{2}}{32} )  \nnn
       + \ltr  [\De u_{\mu} \cm u^{\mu} \cp]\rtr  ( \frac{9}{32} + \frac{N}{64} - \frac{N^{2}}{16} ) 
       + \ltr  [\De u_{\mu} \cp u^{\mu} \cm]\rtr  ( \frac{N^{2}}{64} ) \nnn
              + \ltr \De u_{\mu} \rtr\ltr  [\cp u^{\mu} \cm]\rtr  (  - \frac{1}{4N} - \frac{3N}{32} )
       + \ltr  [\De \cm u^{2} \cp]\rtr  ( \frac{9}{32} + \frac{N}{64} - \frac{5N^{2}}{64} ) \nnn
       + \ltr  [\De u_{\mu} \cm]\rtr\ltr \cp u^{\mu} \rtr  ( \frac{3}{4N} - \frac{N}{96} - \frac{N^{2}}{96} )
       + \ltr  [\De u_{\mu} \cm u^{\mu}]\rtr\ltr \cp \rtr  ( \frac{3}{32} - \frac{1}{8N} + \frac{N}{16} - \frac{N^{2}}{64} ) \nnn
       + \ltr  [\De u_{\mu} \cm u^{\mu} \cp]\rtr  (  - \frac{1}{32} - \frac{1}{N^2} - \frac{3N}{64} - \frac{N^{2}}{32} )
       + \ltr  [\De u_{\mu} \cm \cp u^{\mu}]\rtr  (  - \frac{5}{96} + \frac{5N}{192} ) \nnn
       + \ltr  [\De u_{\mu} \cp]\rtr\ltr \cm u^{\mu} \rtr  ( \frac{1}{8} + \frac{7}{4N} - \frac{11N}{32} )
       + \ltr  [\De u_{\mu} \cp u^{\mu} \cm]\rtr  (  - \frac{5}{32} - \frac{N}{64} + \frac{N^{2}}{64} ) \nnn
       + \ltr  [\De u_{\mu} \cp \cm u^{\mu}]\rtr  (  - \frac{5}{32} - \frac{N}{64} )
       + \ltr  [\De u^{2} \cm]\rtr\ltr \cp \rtr  ( \frac{3}{32} - \frac{1}{4N} + \frac{29N}{96} - \frac{N^{2}}{48} ) \nnn
       + \ltr  [\De u^{2} \cm \cp]\rtr  (  - \frac{7}{48} + \frac{1}{2N^{2}} - \frac{N}{12} + \frac{3N^{2}}{32} )
 + \ltr  [\De \cm u^{2} \cp]\rtr  (  - \frac{19}{96} + \frac{1}{2N^{2}} + \frac{N}{192} - \frac{N^{2}}{64} )
        \nnn
       + \ltr  [\De u^{2} \cp \cm]\rtr  ( \frac{N^{2}}{64} )
       + \ltr  [\De \cm u_{\mu} \cp u^{\mu}]\rtr  ( \frac{29}{96} + \frac{7N}{192} - \frac{N^{2}}{16} ) \nnn
       + \ltr  [\De \cm \cp]\rtr\ltr u^{2} \rtr  ( \frac{1}{16} + \frac{1}{4N} - \frac{N^{2}}{64} )
       + \ltr  [\De \cm \cp u^{2}]\rtr  ( \frac{N}{16} - \frac{3N^{2}}{64} ) \nnn
              + \ltr  [\De u^{2} \cp]\rtr\ltr \cm \rtr  ( \frac{1}{8} - \frac{3}{4N} + \frac{N}{8} )
       + \i\ltr \De u_{\mu} \rtr\ltr \cml{}{\mu} \rtr\ltr \cp \rtr  (  - \frac{1}{16} + \frac{N}{16} ) \nnn
       + \i\ltr \{ \De u_{\mu} \cml{}{\mu} \} \rtr\ltr \cp \rtr  ( \frac{3}{8} - \frac{1}{2N} + \frac{N}{16} - \frac{N^{2}}{16} )
       + \i\ltr \{ \De u_{\mu} \cml{}{\mu} \cp \} \rtr  (  - \frac{3}{8} - \frac{N}{16} + \frac{N^{2}}{8} ) \nnn
       + \i\ltr \{ \De \cml{\mu}{} u^{\mu} \cp \} \rtr  (  - \frac{3}{8} - \frac{N}{16} + \frac{N^{2}}{8} ) 
      + \i\ltr \De u_{\mu} \rtr\ltr  [h^{\mu}_{ \ph \nu}  u^{\nu}  \cp]\rtr  ( \frac{N}{8} )
\nnn
       + \i\ltr \De \cp \rtr\ltr \cpl{\mu}{} u^{\mu} \rtr  ( \frac{1}{8} + \frac{7}{4N} - \frac{3N}{8} - \frac{3N^{2}}{16} )
       + \i\ltr\{\De u_{\mu} \cml{}{\mu}\}\rtr\ltr \cp \rtr  ( \frac{N}{32} - \frac{N^{2}}{32} )
  \nnn
       + \i\ltr\{\De u_{\mu} \cp\}\rtr\ltr \cml{}{\mu} \rtr  (  - \frac{N}{8} )
       + \i\ltr\{\De \cml{\mu}{} u^{\mu} \cp\}\rtr  ( \frac{1}{16} - \frac{N}{32} + \frac{N^{2}}{16} )
        \nnn
       + \i\ltr\{\De u_{\mu} \cml{}{\mu} \cp\}\rtr  ( \frac{1}{16} - \frac{N}{32} + \frac{N^{2}}{16} )
       + \i\ltr  [\De u_{\mu} u_{\nu}  h^{\mu\nu}]\rtr\ltr \cp \rtr  (  - \frac{N}{24} + \frac{N^{2}}{96} ) \nnn
        + \i\ltr  [\De u_{\mu} u_{\nu}  \cp h^{\mu\nu}]\rtr  ( \frac{1}{16} - \frac{N}{32} )
       + \i\ltr  [\De u_{\mu} h^{\mu}_{ \ph \nu}]\rtr\ltr \cp u^{\nu}  \rtr  ( \frac{7N}{48} - \frac{N^{2}}{96} ) \nnn
       + \i\ltr\{\De \cml{\mu}{} \cp u^{\mu}\}\rtr  (  - \frac{1}{8} + \frac{N}{16} )
       + \i\ltr  [\De u_{\mu} \cp]\rtr\ltr h^{\mu}_{ \ph \nu}  u^{\nu}  \rtr  (  - \frac{1}{N} 
       + \frac{N}{4} ) \nnn
       + \i\ltr  [\De u_{\mu} \cp u_{\nu}  h^{\mu\nu}]\rtr  (  - \frac{1}{48} + \frac{N}{96} )  
       + \i\ltr  [\De \cp u_{\mu} u_{\nu}  h^{\mu\nu}]\rtr  ( \frac{1}{24} - \frac{N}{48} - \frac{N^{2}}{16} ) \nnn   
       + \i\ltr  [\De u_{\mu} \cp h^{\mu}_{ \ph \nu} u^{\nu} ]\rtr  (  - \frac{5}{48} + \frac{5N}{96} )
       + \i\ltr  [\De \cp u_{\mu} h^{\mu}_{ \ph \nu} u^{\nu} ]\rtr  ( \frac{7}{16} + \frac{N}{32} - \frac{N^{2}}{16} )
       \no
\eea
\bea
& & 
       + \i\ltr  [\De \cp h_{\mu\nu} ]\rtr\ltr u^{\mu} u^{\nu}  \rtr  ( \frac{1}{2N} - \frac{3N}{32} )
       + \i\ltr  [\De \cp h_{\mu\nu}  u^{\mu} u^{\nu} ]\rtr  ( \frac{1}{12} - \frac{N}{24} - \frac{N^{2}}{32} ) \nnn 
        + \i\ltr \De u_{\mu} \rtr\ltr \cpl{}{\mu} \rtr\ltr \cm \rtr  ( \frac{1}{8} + \frac{1}{2N^{2}} 	- \frac{3}{8N} - \frac{N}{16})  
       + \i\ltr \De u_{\mu} \rtr\ltr \cpl{}{\mu} \cm \rtr  ( \frac{1}{2} - \frac{3}{4N} + \frac{5N}{16} 	- \frac{N^{2}}{16} ) \nnn
      + \i\ltr \De \cpl{\mu}{} \rtr\ltr \cm u^{\mu} \rtr  ( \frac{1}{8} - \frac{1}{4N} + \frac{3N}{16} )
       + \i\ltr \De \cm \rtr\ltr \cpl{\mu}{} u^{\mu} \rtr  ( \frac{1}{8} + \frac{3}{4N} + \frac{N}{16} )     \nnn
       + \i\ltr\{\De u_{\mu} \cm\}\rtr\ltr \cpl{}{\mu} \rtr  ( \frac{1}{2N} + \frac{N}{32} )
       + \i\ltr\{\De u_{\mu} \cm \cpl{}{\mu}\}\rtr  (  - \frac{1}{4} - \frac{3N}{32} ) \nnn
       + \i\ltr\{\De \cm u_{\mu} \cpl{}{\mu}\}\rtr  (  - \frac{1}{4} - \frac{N}{16} - \frac{N^{2}}{32} )   
       + \i\ltr\{\De \cm \cpl{\mu}{} u^{\mu}\}\rtr  (  - \frac{5}{8} - \frac{3N}{32} + \frac{N^{2}}{16} ) \nnn
       + \ltr \De u_{\mu} \rtr\ltr \cpl{\ph \nu\}}{\{\mu} u^{\nu}  \rtr  (  - \frac{1}{3N} + \frac{N}{8} )
       + \ltr \De u_{\mu} \rtr\ltr \cpl{\nu}{\nu}  u^{\mu} \rtr  (  - \frac{4}{3N} + \frac{3N}{8} ) \nnn
       + \ltr \De u_{\mu} u_{\nu}  \rtr\ltr \cpl{}{\{\mu\nu\}} \rtr  ( \frac{1}{6N} + \frac{N}{72} )
       + \ltr \De u_{\mu} \cpl{\nu}{\nu} u^{\mu} \rtr  (  - \frac{1}{36} - \frac{N^{2}}{72} )
       + \ltr \De \cpl{\mu}{\mu} \rtr\ltr u^{2} \rtr  (  - \frac{2}{3N} - \frac{N}{72} ) \nnn
       + \i\ltr\{\De u_{\mu} \cpl{}{\mu}\}\rtr\ltr \cm \rtr  ( \frac{1}{8} - \frac{1}{8N} + \frac{3N}{32} )      + \ltr \De \cpl{\mu\nu}{} \rtr\ltr u^{\mu} u^{\nu}  \rtr  (  - \frac{1}{3N} + \frac{N}{18} )\nnn
       + \ltr \De u^{2} \rtr\ltr \cpl{\mu}{\mu} \rtr  ( \frac{2}{3N} + \frac{11N}{36} )
       + \ltr\{\De u_{\mu} u_{\nu}  \cpl{}{\{\mu\nu\}}\}\rtr  ( \frac{1}{72} - \frac{N^{2}}{144} )  
	+ \ltr \De \cm \rtr\ltr \cm u^{2} \rtr  (  - \frac{1}{8} - \frac{3}{2N} ) \nnn
       + \ltr\{\De u^{2} \cpl{\mu}{\mu}\}\rtr  (  - \frac{23}{72} + \frac{23N^{2}}{144} )  
       + \i\ltr  [\De \cm u_{\mu} \cml{}{\mu}]\rtr ( \frac{N^{2}}{32} )
       + \ltr\{\De u_{\mu} \cpl{\ph\nu}{\mu} u^{\nu} \}\rtr  (  - \frac{7}{36} + \frac{N^{2}}{36} )
       \nnn
       + \i\ltr  [\De u_{\mu} \cml{}{\mu} \cm]\rtr  (  - \frac{N^{2}}{32} )
       + \i\ltr  [\De u_{\mu} \cm]\rtr\ltr \cml{}{\mu} \rtr  ( \frac{N}{16} )  
       + \i\ltr  [\De \cm u_{\mu} \cml{}{\mu}]\rtr  (  - \frac{1}{4} + \frac{N^{2}}{16} )\nnn
       + \i\ltr  [\De \cm \cpl{\mu}{} u^{\mu}]\rtr  (  - \frac{1}{4} + \frac{N^{2}}{16} )
       + \ltr \De u_{\mu} \rtr\ltr \cm u^{\mu} \rtr\ltr \cm \rtr  (  - \frac{1}{8} - \frac{1}{2N^{2}} + \frac{3}{8N} + \frac{N}{16}) \nnn
       + \ltr \De u_{\mu} \rtr\ltr \cm \cm u^{\mu} \rtr  (  - \frac{1}{2} + \frac{1}{N} - \frac{N}{4} + \frac{N^{2}}{16} )
       + \ltr \De u_{\mu} \cm \rtr\ltr \cm u^{\mu} \rtr  (  - \frac{1}{16} + \frac{1}{8N} - \frac{3N}{32} )\nnn 
       + \ltr \De u^{2} \rtr\ltr \cm \cm \rtr  (  - \frac{1}{8N} - \frac{N}{32} ) 
        + \ltr \De \cm \cm \rtr\ltr u^{2} \rtr  (  - \frac{5}{8N} + \frac{3N}{32} )
       \nnn
       + \ltr \De \cm u_{\mu} \rtr\ltr \cm u^{\mu} \rtr  (  - \frac{1}{16} + \frac{1}{8N} - \frac{3N}{32} )
       + \ltr \De \cm u^{2} \cm \rtr  (   \frac{5}{24} + \frac{1}{N^2} + \frac{N}{32} 
       +\frac{7 N^{2}}{96} )  \nnn
       + \ltr\{\De u_{\mu} \cm\}\rtr\ltr \cm u^{\mu} \rtr  ( \frac{1}{4N} - \frac{3N}{32} )
       + \ltr \De u_{\mu} \cm u^{\mu} \rtr\ltr \cm \rtr  (  - \frac{1}{8} + \frac{1}{8N} - \frac{N}{16} ) \nnn
       + \ltr\{\De u_{\mu} \cm u^{\mu} \cm\}\rtr  (   \frac{13}{48} - \frac{1}{N^2} + \frac{3N}{32} - \frac{N^{2}}{96} )  
       + \ltr\De u_{\mu} \cm \cm u^{\mu} \rtr  ( \frac{5}{24} + \frac{N}{8} ) \nnn
       + \ltr\{\De u^{2} \cm\}\rtr\ltr \cm \rtr  (  - \frac{1}{16} - \frac{5}{16N} - \frac{N}{32} )
       + \ltr\{\De u^{2} \cm \cm\}\rtr  ( \frac{1}{3} + \frac{1}{2N^{2}} + \frac{N}{16} - \frac{N^{2}}{96} )  \nnn
       + \ltr \De u_{\mu} \rtr\ltr h_{\nu\rho}  h^{\nu\rho}  u^{\mu} \rtr  ( \frac{3N}{16} )
       + \ltr \De u_{\mu} \rtr\ltr\{h^{\mu}_{ \ph \nu} h^{\nu}_{\ph \rho}  u^{\rho}\}\rtr  ( \frac{N}{16} )
        + \ltr \De u_{\mu} u_{\nu} \rtr\ltr h^{\mu}_{ \ph \rho} h^{\nu\rho} \rtr  ( \frac{N}{36} )
\nnn  
       + \ltr \De u_{\mu} u_{\nu} \rtr\ltr h^{\nu}_{\ph \rho} h^{\mu\rho} \rtr  (  - \frac{N}{72} )
        + \ltr \De u_{\mu} h^{\mu}_{ \ph \nu} h^{\nu}_{ \ph \rho}  u^{\rho} \rtr  (  - \frac{N^{2}}{36} )
       + \ltr \De u_{\mu} h_{\nu\rho}  h^{\nu\rho}  u^{\mu} \rtr  (  - \frac{N^{2}}{144} ) \nnn 
       + \ltr \De h_{\mu\nu}  \rtr\ltr h^{\mu\nu}  u^{2} \rtr  ( \frac{N}{72} )
       + \ltr \De h_{\mu\nu}  \rtr\ltr\{h^{\nu}_{ \ph \rho}  u^{\rho} u^{\mu}\}\rtr  (  - \frac{N}{36} )
       + \ltr \De h_{\mu\nu}  u^{\mu} u_{\rho} h^{\nu\rho}  \rtr  (  - \frac{1}{12} )
       \no
\eea
\bea
& &  
       + \ltr \De h_{\mu\nu}  u_{\rho} u^{\mu} h^{\nu\rho}  \rtr  (  - \frac{1}{12} )
       + \ltr \De h_{\mu\nu}  h^{\mu\nu}  \rtr\ltr u^{2} \rtr  (  - \frac{N}{144} )
       + \ltr \De h_{\mu\nu}  h^{\mu}_{\ph \rho} \rtr\ltr u^{\nu}  u^{\rho} \rtr  ( \frac{N}{36} ) \nnn
       + \ltr \De h_{\mu\nu}  u^{2} h^{\mu\nu}  \rtr  (  - \frac{1}{3} )
       + \ltr \De u^{2} \rtr\ltr h_{\mu\nu}  h^{\mu\nu}  \rtr  ( \frac{11N}{72} )
       + \ltr\{\De u_{\mu} u_{\nu} h^{\mu}_{\ph \rho} h^{\nu\rho}\}\rtr  ( \frac{1}{24} ) \nnn
                  + \ltr\{\De u_{\mu} u_{\nu} h^{\nu}_{\ph \rho} h^{\mu\rho}\}\rtr  ( \frac{1}{24} - \frac{N^{2}}{144} )
       + \ltr\{\De u_{\mu} h^{\mu}_{ \ph \nu}\}\rtr\ltr h^{\nu}_{\ph \rho}  u^{\rho} \rtr  (  - \frac{N}{48} )
       + \ltr\{\De u_{\mu} h^{\mu}_{ \ph \nu} h^{\nu}_{\ph \rho}  u^{\rho}\}\rtr  ( \frac{N^{2}}{36} ) \nnn  
       + \ltr\{\De u_{\mu} h_{\nu\rho} \}\rtr\ltr h^{\mu\rho} u^{\nu}  \rtr  (  - \frac{N}{48} )
       + \ltr\{\De u_{\mu} h_{\nu\rho} \}\rtr\ltr h^{\nu\rho}  u^{\mu} \rtr  (  - \frac{N}{48} )
       + \ltr\{\De u^{2} h_{\mu\nu}  h^{\mu\nu} \}\rtr  ( \frac{1}{6} + \frac{23N^{2}}{288} )  \nnn
       + \i\ltr \De u_{\mu} \rtr\ltr  [\cpl{\nu}{} u^{\mu} u^{\nu}]\rtr  (  - \frac{N}{48} )
       + \i\ltr  [\De u_{\mu} u_{\nu} ]\rtr\ltr \cpl{}{\mu} u^{\nu} \rtr  (  - \frac{N}{48} )
       + \i\ltr  [\De u_{\mu} u_{\nu} u^{\mu} \cpl{}{\nu}]\rtr  (  - \frac{5}{24} ) \nnn  
       + \i\ltr  [\De u_{\mu} u_{\nu} \cpl{}{\mu} u^{\nu}]\rtr  ( \frac{1}{8} )
       + \i\ltr  [\De u_{\mu} u_{\nu} \cpl{}{\nu} u^{\mu}]\rtr  ( \frac{1}{8} )
       + \i\ltr  [\De u_{\mu} \cpl{}{\mu}]\rtr\ltr u^{2} \rtr  ( \frac{1}{4N} - \frac{N}{12} ) \nnn  
       + \i\ltr  [\De u_{\mu} \cpl{\nu}{} ]\rtr\ltr u^{\mu} u^{\nu}  \rtr  ( \frac{1}{2N} + \frac{N}{48} )
       + \i\ltr  [\De u_{\mu} u^{2}]\rtr\ltr \cpl{}{\mu} \rtr  ( \frac{11N}{48} )
       + \i\ltr  [\De u_{\mu} u^{2} \cpl{}{\mu}]\rtr  ( \frac{1}{6} + \frac{N^{2}}{48} )  \nnn
       + \i\ltr  [\De u^{2} u_{\mu} \cpl{}{\mu}]\rtr  ( \frac{1}{6} - \frac{7N^{2}}{48} )
       + \i\ltr  [\De u^{2} \cpl{\mu}{} u^{\mu}]\rtr  ( \frac{1}{8} - \frac{N^{2}}{16} )
        + \i\ltr \De u_{\mu} \rtr\ltr u^{\mu} u_{\nu} \rtr\ltr \cml{}{\nu} \rtr  (  - \frac{1}{3} ) \nnn  
       + \i\ltr \De u_{\mu} \rtr\ltr \cml{}{\mu} u^{2} \rtr  (  - \frac{2}{3N} + \frac{19N}{72} )
       + \i\ltr \De u_{\mu} \rtr\ltr \{ \cml{\nu}{} u^{\mu} u^{\nu} \} \rtr  (  - \frac{5}{3N} + \frac{11N}{16} ) \nnn
       + \i\ltr \{ \De u_{\mu} u_{\nu} \} \rtr\ltr \cml{}{\mu} u^{\nu} \rtr  ( \frac{1}{3N} + \frac{N}{144} )
       + \ltr \De u_{\mu} \rtr\ltr\{\cp u^{\mu} u^{2}\}\rtr  (  - \frac{3}{4N} - \frac{5N}{32} + \frac{N^{2}}{32} )
\nnn
       + \i\ltr \De u_{\mu} u_{\nu} u^{\mu} \rtr\ltr \cml{}{\nu} \rtr  ( \frac{N}{72} )  
       + \i\ltr \De u_{\mu} u_{\nu} \cml{}{\nu} u^{\mu} \rtr  (  - \frac{1}{36} - \frac{N^{2}}{72} )
       + \i\ltr \De u_{\mu} \cml{}{\mu} \rtr\ltr u^{2} \rtr  (  - \frac{2}{3N} - \frac{N}{72} ) \nnn
       + \i\ltr \De u_{\mu} \cml{\nu}{} \rtr\ltr u^{\mu} u^{\nu} \rtr  (  - \frac{1}{3N} + \frac{N}{18} )  
       + \i\ltr \De u_{\mu} \cml{\nu}{} u^{\nu} u^{\mu} \rtr  (  - \frac{1}{36} - \frac{N^{2}}{72} )
       + \i\ltr \De \cpl{\mu}{} \rtr\ltr u^{2} u^{\mu} \rtr  ( \frac{N}{72} ) \nnn
       + \i\ltr \De \cpl{\mu}{} u^{\mu} \rtr\ltr u^{2} \rtr  (  - \frac{2}{3N} - \frac{N}{72} )  
       + \i\ltr \De \cpl{\mu}{} u_{\nu} \rtr\ltr u^{\mu} u^{\nu} \rtr  (  - \frac{1}{3N} + \frac{N}{18} ) \nnn
       + \i\ltr\{\De u_{\mu} u_{\nu} u^{\mu} \cml{}{\nu}\}\rtr  ( \frac{1}{72} - \frac{N^{2}}{288} )  
       + \i\ltr\{\De u_{\mu} u_{\nu} \cml{}{\mu} u^{\nu}\}\rtr  (  - \frac{1}{12} - \frac{N^{2}}{288} ) \nnn
              + \i\ltr \De u^{2} \rtr\ltr \cpl{\mu}{} u^{\mu} \rtr  ( \frac{4}{3N} + \frac{11N}{12} )
       + \i\ltr\{\De u_{\mu} u_{\nu} \cml{}{\nu} u^{\mu}\}\rtr  ( \frac{1}{72} - \frac{5N^{2}}{288} ) \nnn
       + \i\ltr\{\De u_{\mu} \cml{}{\mu}\}\rtr\ltr u^{2} \rtr  (  - \frac{N}{144} )  
       + \i\ltr\{\De u_{\mu} \cml{}{\mu} u^{2}\}\rtr  (  - \frac{7}{72} + \frac{N^{2}}{72} )
       + \i\ltr\{\De u_{\mu} \cml{\nu}{}\}\rtr\ltr u^{\mu} u^{\nu} \rtr  (  - \frac{N}{48} ) \nnn
       + \i\ltr\{\De u_{\mu} \cml{\nu}{} u^{\mu} u^{\nu}\}\rtr  (  - \frac{7}{72} + \frac{N^{2}}{72} )  
        + \ltr \De u_{\mu} \rtr\ltr \cp u^{\mu} u^{2} \rtr  ( \frac{7}{12N} - \frac{83N}{288} )
      \nnn
       + \i\ltr\{\De u^{2} u_{\mu} \cml{}{\mu}\}\rtr  (  - \frac{11}{72} + \frac{23N^{2}}{96} )  
       + \i\ltr\{\De u^{2} \cpl{\mu}{} u^{\mu}\}\rtr  (  - \frac{1}{4} + \frac{77N^{2}}{288} )
       \nnn
         + \i\ltr\{\De u_{\mu} u^{2} \cml{}{\mu}\}\rtr  (  - \frac{23}{72} - \frac{N^{2}}{96} )
       + \ltr \De u_{\mu} \rtr\ltr u^{\mu} u_{\nu} \rtr\ltr \cp u^{\nu} \rtr  (  - \frac{5}{12} + \frac{N}{8} ) \nnn
       + \ltr \De u_{\mu} \rtr\ltr u^{2} \rtr\ltr \cp u^{\mu} \rtr  ( \frac{1}{4} + \frac{N}{8} )  
       + \ltr \De u_{\mu} \rtr\ltr u^{2} u^{\mu} \rtr\ltr \cp \rtr  (  - \frac{3}{8} + \frac{N}{8} )
       + \i\ltr\{\De u_{\mu} u^{2}\}\rtr\ltr \cml{}{\mu} \rtr  (  - \frac{23N}{144} )
       \nnn
       + \ltr \De u_{\mu} \rtr\ltr \cp u_{\nu} u^{\mu} u^{\nu} \rtr  ( \frac{7}{3N} - \frac{9N}{16} )  
       + \ltr \De u_{\mu} \rtr\ltr \cp u^{2} u^{\mu} \rtr  ( \frac{7}{12N} - \frac{83N}{288} )
       \no
\eea
\bea
& & 
       + \ltr \De u_{\mu} u_{\nu} \rtr\ltr \cp u^{\mu} u^{\nu} \rtr  (  - \frac{1}{6N} - \frac{11N}{144} )  
       + \ltr \De u_{\mu} u_{\nu} \rtr\ltr \cp u^{\nu} u^{\mu} \rtr  (  - \frac{1}{6N} - \frac{5N}{144} )
       \nnn
       + \ltr \De u_{\mu} u_{\nu} u^{\mu} u^{\nu} \rtr\ltr \cp \rtr  ( \frac{N}{24} - \frac{N^{2}}{16} )  
       + \ltr \De u_{\mu} u_{\nu} \cp \rtr\ltr u^{\mu} u^{\nu} \rtr  ( \frac{1}{12N} - \frac{N}{72} ) \nnn
       + \ltr \De u_{\mu} u_{\nu} u^{\mu} \rtr\ltr \cp u^{\nu} \rtr  (  - \frac{23N}{144} )
       + \ltr \De u_{\mu} u_{\nu} \cp u^{\mu} u^{\nu} \rtr  (  - \frac{3}{16} + \frac{N^{2}}{96} ) \nnn
       + \ltr \De u_{\mu} u_{\nu} \cp u^{\nu} u^{\mu} \rtr  ( \frac{83}{144} - \frac{N}{8} + \frac{5N^{2}}{288} )  
+ \ltr \De u_{\mu} \cp u_{\nu} \rtr\ltr u^{\mu} u^{\nu} \rtr  ( \frac{1}{6N} - \frac{13N}{144} )
       \nnn
      + \ltr \De u_{\mu} \cp \rtr\ltr u^{2} u^{\mu} \rtr  (  - \frac{N}{144} )  
       + \ltr \De u_{\mu} \cp u^{\mu} \rtr\ltr u^{2} \rtr  ( \frac{13}{12N} + \frac{N}{144} ) 
       + \ltr \De u_{\mu} u^{2} u^{\mu} \rtr\ltr \cp \rtr  ( \frac{N}{48} + \frac{N^{2}}{16} )       
       \nnn 
              + \ltr \De \{ u_{\mu} \cp u^{2} u^{\mu} \} \rtr  ( \frac{1}{144} + \frac{N^{2}}{288} )  
       + \ltr \De u^{2} \rtr\ltr u^{2} \rtr\ltr \cp \rtr  (  - \frac{1}{16} + \frac{N}{16} )\nnn
              + \ltr \De u^{2} \rtr\ltr \cp u^{2} \rtr  (  - \frac{2}{3N} - \frac{13N}{36} + \frac{N^{2}}{16} )
       + \ltr \De u^{2} u^{2} \rtr\ltr \cp \rtr  (  - \frac{5N}{16} + \frac{N^{2}}{16} ) \nnn 
       + \ltr \De u^{2} \cp \rtr\ltr u^{2} \rtr  ( \frac{1}{6N} + \frac{N}{288} )
       + \ltr \De u^{2} \cp u^{2} \rtr  (  - \frac{3}{16} - \frac{N}{16} - \frac{11N^{2}}{96} ) \nnn
       + \ltr \De \cp \rtr\ltr u_{\mu} u_{\nu} u^{\mu} u^{\nu} \rtr  (  - \frac{N}{6} - \frac{N^{2}}{16} )  
       + \ltr \De \cp \rtr\ltr u^{2} \rtr\ltr u^{2} \rtr  ( \frac{3}{16} + \frac{N}{32} )
       + \ltr \De \cp \rtr\ltr u^{2} u^{2} \rtr  ( \frac{5N}{12} + \frac{N^{2}}{16} ) \nnn
       + \ltr \De \cp u_{\mu} \rtr\ltr u^{2} u^{\mu} \rtr  (  - \frac{N}{144} )  
       + \i\ltr \De u_{\mu} \rtr\ltr h^{\mu}_{ \ph \nu} u^{\nu}  \rtr\ltr \cm \rtr  ( \frac{1}{8} - \frac{N}{16} )
       + \ltr \De u_{\mu} \rtr\ltr h^{\mu}_{\ph \nu\rho} u^{\nu}  u^{\rho} \rtr  (  - \frac{N}{8} ) \nnn
       + \i\ltr \De u_{\mu} \rtr\ltr\{h^{\mu}_{ \ph \nu} u^{\nu}  \cm\}\rtr  ( \frac{N}{16} - \frac{N^{2}}{32} )  
       + \ltr \De u_{\mu} \rtr\ltr h^{\mu}_{\ph \nu\rho} u^{\nu}  u^{\rho} \rtr  ( \frac{N}{4} )
       + \ltr \De u_{\mu} \rtr\ltr\{h_{\nu \ph \rho}^{\ph \mu} u^{\nu}  u^{\rho}\}\rtr  ( \frac{5N}{144} ) \nnn
       + \i\ltr \De u_{\mu} u_{\nu}  \rtr\ltr h^{\mu\nu} \cm \rtr  (  - \frac{3N}{16} )  
       + \ltr \De u_{\mu} u_{\nu}  \rtr\ltr h^{\mu\nu}_{\ph \ph \rho} u^{\rho} \rtr  (  - \frac{N}{144} )
       + \ltr \De u_{\mu} u_{\nu}  \rtr\ltr h^{\nu\mu}_{ \ph \ph \rho} u^{\rho} \rtr  (  - \frac{N}{144} ) \nnn
       + \ltr \De u_{\mu} u_{\nu}  \rtr\ltr h_{\rho}^{\ph \mu\nu} u^{\rho} \rtr  (  - \frac{N}{24} )
       + \i\ltr \De u_{\mu} h^{\mu}_{ \ph \nu} u^{\nu}  \rtr\ltr \cm \rtr  (   \frac{N}{8} )
       + \i\ltr \De h_{\mu\nu}  \rtr\ltr \cm u^{\mu} u^{\nu}  \rtr  (  - \frac{3N}{16} ) \nnn
       + \ltr\{\De u_{\mu} u_{\nu}  u_{\rho} h^{\mu\nu\rho}\}\rtr  ( \frac{1}{24} )
       + \ltr\{\De u_{\mu} u_{\nu}  u_{\rho} h^{\nu\mu\rho}\}\rtr  (  - \frac{1}{24} - \frac{N^{2}}{144} )
       + \ltr\{\De u_{\mu} u_{\nu}  u_{\rho} h^{\rho\mu\nu}\}\rtr  (  - \frac{1}{12} ) \nnn
       + \i\ltr\{\De u_{\mu} u_{\nu}  h^{\mu\nu}\}\rtr\ltr \cm \rtr  ( \frac{N}{32} )  
       + \ltr\{\De u_{\mu} u_{\nu}  h^{\mu\nu}_{\ph \ph \rho} u^{\rho}\}\rtr  ( \frac{1}{24} + \frac{N^{2}}{48} )
       + \ltr\{\De u_{\mu} u_{\nu}  h^{\nu\mu}_{\ph \ph \rho} u^{\rho}\}\rtr  ( \frac{1}{24} ) \nnn
       + \i\ltr\{\De u_{\mu} u_{\nu}  \cm h^{\mu\nu}\}\rtr  (  - \frac{1}{8} + \frac{3N}{32} )
       + \ltr\{\De u_{\mu} h^{\mu}_{\ph \nu\rho}\}\rtr\ltr u^{\nu}  u^{\rho} \rtr  (  - \frac{N}{48} ) \nnn
       + \ltr\{\De u_{\mu} h_{\nu \ph \rho }^{\ph \mu}\}\rtr \ltr u^{\nu}  u^{\rho} \rtr
       ( \frac{N}{144} )
       + \i\ltr \De \cm \rtr\ltr h_{\mu\nu}  u^{\mu} u^{\nu}  \rtr  ( \frac{N}{16} )
	- \ltr \De h_{\mu\nu\rho}  \rtr\ltr u^{\mu} u^{\nu}  u^{\rho} \rtr  ( \frac{N}{18} ) \nnn
       + \i\ltr\{\De u_{\mu} \cm u_{\nu}  h^{\mu\nu}\}\rtr  ( \frac{1}{48} - \frac{N}{32} )
       + \i\ltr\{\De u_{\mu} \cm h^{\mu}_{ \ph \nu} u^{\nu} \}\rtr  (  - \frac{1}{48} + \frac{N}{32} ) \nnn
       + \i\ltr\{\De \cm u_{\mu} u_{\nu}  h^{\mu\nu}\}\rtr  ( \frac{1}{6} - \frac{N^{2}}{96} )  
       + \i\ltr\{\De \cm u_{\mu} h^{\mu}_{ \ph \nu} u^{\nu} \}\rtr  (  - \frac{3}{16} - \frac{3N}{32} + \frac{N^{2}}{48} )
       \nnn
       + \i\ltr\{\De \cm h_{\mu\nu} \}\rtr\ltr u^{\mu} u^{\nu}  \rtr  (  - \frac{N}{32} )
       + \i\ltr\{\De \cm h_{\mu\nu}  u^{\mu} u^{\nu} \}\rtr  ( \frac{7}{48} + \frac{N^{2}}{48} )
\eea
\newpage
\noindent $\ltr \cdot \rtr$ denotes the trace in $SU(N)$. We calculated the generic case with $N$ flavors in order to make the above counterterm also usable for the quenched case          \cite{Sharpe:1992ft,Colangelo:1997ch}. The calculation was performed in euclidean space-time,
but above we provide the result transformed back to Minkowski space.
Furthermore we used the notation:
\begin{alignat}{2}
 [ A_{1}...A_{n}] & := A_{1}A_{2}...A_{n} - A_{n}A_{n-1}...A_{1} \; ; \quad &
 \{ A_{1}...A_{n} \} & := A_{1}A_{2}...A_{n} + A_{n}A_{n-1}...A_{1} \no \fs
\end{alignat}
The above expression corresponds only to the part of $\mA_{2}^{(2)}$,
which does not involve any external vector sources and can contribute to processes involving
four pseudo Goldstone fields, i.e. can be used for  the (physical) \ktpp and \ktppp amplitudes.
For obvious reasons we do not provide a minimal basis for $\mL^{(2)}_{\tiny{w}}$.
However, for the part of $\mA_{2}^{(2)}$ shown here, all operators are linearly independent:
One can use Cayley-Hamilton relations (CHR), the Bianchi identities and partial
integrations to (potentially) reduce the number of operators and find a basis.
Yet, the CHR are not usable since we worked in general $SU(N)$, the Bianchi identities
involve vector-sources which were neglected, and partial integrations cannot be employed
 since it would necessarily generate an operator involving $\Dem$ \footnote{In a full basis
 it would however be preferable to trade operators within $\tilde{\mA}^{(2)}_{2}$ for operators
 involving $\Dem$.}.
 \\
We do not to present the full result, including also the operators containing $\Dem$ and $\fipm$, here, since it is approximately a factor four times larger. It can however be obtained from the author. \\
\setcounter{equation}{0}
\section{Notation/Definitions}
\label{s:nd}
The notation is as follows:
\begin{alignat}{2}
 U   & :=  u^{2} := \mexp ( \frac{\i \sqrt{2}\phi}{F})  \sem \quad &
   U & \longrightarrow g_{R} U g_{L}^{\dagger} \no \co
\end{alignat}
with $\phi$, defined in Eq. (\ref{eq:phid}), representing the pseudo Goldstone bosons and the arrow showing 
the response of $U$ to a chiral transformation $(g_{L},g_{R}) \in \big(SU(3)_{L},SU(3)_{R}\big)$.\\
The building blocks used are:
\bea
 \De & := &   u \lambda_{6} u^{\dagger} \con
 \um & := & \i \big( u^{\dagger}(\pa\dom - \i r\dom )u - u(\pa\dom - \i l\dom) u^{\dagger} \big) \con
  \chi_{_{\pm}} & := & u^{\dagger} \chi u^{\dagger} \pm u \chi^{\dagger} u \con
 \Gamma\dom & := & \frac{1}{2} \big( u^{\dagger}(\pa\dom - \i r\dom )u
                                  + u(\pa\dom - \i l\dom) u^{\dagger} \big) \con
 \Dm X & := & \pa\dom X  + [\Gamma\dom, X ] \con
 \Dem & := & \Dm \De + \frac{\i}{2} \{ \De, \um \} \co \label{eq:nocl} \\
 \umn & := & \Dm \un \con
\hmn & := & \Dm \un + \Dn \um \con
 h_{\mu\nu\rho} & := & \Dm h_{\nu\rho} \con
 \chi_{_{\pm},\mu} & := & \Dm\chi_{_{\pm}} -\frac{\i}{2}\{\chi_{\mp},\um \} \con
 \chi_{_{\pm}\mu} & := & \Dm\chi_{_{\pm}} \con
 \chi_{_{\pm}\{\mu,\nu\}} & := & \{\Dm,\Dn\} \chi_{_{\pm}} \fs \no
\eea
All of these transform like:
\be
 X \to h(\vph,g) X h ^{\dagger}(\vph,g) \co \no
\ee
where $h(\vph,g)$ is the compensating $SU(N)_{V}$ transformation defined:
\be
u \to g_{R} u h^{\dagger}(\vph,g) = h(\vph,g) u g_{L}^{\dagger} \co \no
\ee
with the exception of $\Dem$ which transforms like $\Dem \to g_{L} \Dem g_{L}^{\dagger}$.
In addition we used notation that derives from the above ($u^{2} := u_{\mu}u^{\mu},\;h_{\mu\nu\rho\sigma} := \Dm\Dn h_{\rho\sigma}$ etc.). All calculations were performed with FORM 3.1
\cite{Vermaseren:2000nd}.\\
Furthermore, we used:
 \begin{alignat}{3}
 \eps & := 4-d \sem \quad & \hat{N} & := (4\pi)^{-2} \sem \quad & \Lambda  & :=   \frac{\hat{N}}{\eps} \fs 
\end{alignat}
\newpage
\setcounter{equation}{0}
\section{Expansion of the building blocks in the $\xi$-fields}
\label{app:bbexp}
Below we provide the expansion of the building blocks given in appendix \ref{s:nd} in terms of 
the quantum fluctuation fields $\xi$. 
The building blocks on the RHS are evaluated at the EOM (expressed by the bar). 
\bea
 \De & = & \Deb - \frac{\i}{2} [\Deb,\xi] - \frac{1}{8} [[\Deb,\xi],\xi] +
 \frac{\i}{48}[[[\Deb,\xi],\xi],\xi] +\mO(\xi^{4}) \con 
 \um & = & \umb -\xim -  \frac{1}{8}[[\umb,\xi],\xi] + \frac{1}{24} [[\xim,\xi],\xi] +\mO(\xi^{4}) \con 
 \Gamma_{\mu} & = & \bar{\Gamma}_{\mu} 
 + \frac{1}{4} [\umb,\xi] - \frac{1}{8} [\xim,\xi]
  - \frac{1}{96} [[[\umb,\xi],\xi],\xi] +\mO(\xi^{4}) \con 
 \cipm & = & \cipmb
 - \frac{\i}{2} \{ \cimpb, \xi \} - \frac{1}{8} \{\{\cipmb,\xi\},\xi\}
 + \frac{\i}{48} \{\{\{\cimpb,\xi\},\xi\},\xi\} +\mO(\xi^{4}) \con 
 \fipm & = & \fipmb -\frac{\i}{2} [\fimpb,\xi] - \frac{1}{8} [[\fipmb,\xi],\xi]
 +\frac{\i}{48} [[[\fimpb,\xi],\xi],\xi] + \mO(\xi^{4}) \con 
 \Dem & = & \Demb - \frac{\i}{2} [\Demb,\xi] - \frac{1}{8} [[\Demb,\xi],\xi]
 + \frac{\i}{48} [[[\Demb,\xi],\xi],\xi] + \mO(\xi^{4}) \con 
 \dcihpm & = & \dcihpmb - \frac{\i}{2} \{ \dcihmpb,\xi\} 
 - \frac{1}{8} [[\dcihpmb,\xi],\xi] + \mO(\xi^{3}) \con 
 \umn & = & \umnb - \xi_{\mu\nu} + \frac{1}{4}[[\umb,\xi],\unb] - \frac{1}{8} [[\umnb,\xi],\xi] \con 
 & & -\frac{1}{4} [[\umb,\xi],\xin] - \frac{1}{4} [[\unb,\xi],\xim] + \mO(\xi^{3}) \con 
 \hat{X} & = & -\xi_{\mu\mu} + \frac{1}{4} [[\umb,\xi],\umb] 
 + \frac{1}{4} \{\cpb,\xi\}
 - \frac{1}{2N} \ltr \cpb \xi \rtr \con 
 & & - \frac{1}{2} [[\umb,\xi],\xim] - \frac{1}{4} \xi\cmb\xi + \frac{1}{4N} \ltr \cmb\xi^{2} \rtr + \mO(\xi^{3}) \fs 
 \label{eq:xiexp}
\eea
We use the conventions $u = \bar{u}\exp(\i\xi/2)$\;,\;
$\xi = \sum \lambda_{a} \xi_{a}$\;,\; $\ltr \lambda_{a} \lambda_{b} \rtr = 2\delta_{ab}$ and the notation
$\xi_{\mu} := \nabla_{\mu}\xi$\;,\; $\xi_{\mu\nu} := \nabla_{\mu}\nabla_{\nu} \xi$. \\
\newpage
\bibliography{list.bib}
\end{document}